# What Do Citation Counts Measure? An Updated Review of Studies on Citations in Scientific Documents Published between 2006 and 2018

**Iman Tahamtan**[1] & Lutz Bornmann[2]

[1] School of Information Sciences, College of Communication and Information, University of Tennessee, Knoxville, TN, USA

**Corresponding author: Iman Tahamtan:** Email: tahamtan@vols.utk.edu

[2] Administrative Headquarters of the Max Planck Society, Division for Science and Innovation Studies, Hofgartenstr. 8, 80539 Munich, Germany

Email: bornmann@gv.mpg.de

## Abstract

The purpose of this paper is to update the review of Bornmann and Daniel (2008) presenting a narrative review of studies on citations in scientific documents. The current review covers 41 studies published between 2006 and 2018. Bornmann and Daniel (2008) focused on earlier years. The current review describes the (new) studies on citation content and context analyses as well as the studies that explore the citation motivation of scholars through surveys or interviews. One focus in this paper is on the technical developments in the last decade, such as the richer meta-data available and machine-readable formats of scientific papers. These developments have resulted in citation context analyses of large datasets in comprehensive studies (which was not possible previously). Many studies in recent years have used computational and machine learning techniques to determine citation functions and polarities, some of which have attempted to overcome the methodological weaknesses of previous studies. The automated recognition of citation functions seems to have the potential to greatly enhance citation indices and information retrieval capabilities. Our review of the empirical studies demonstrates that a paper may be cited for very different scientific and non-scientific reasons. This result accords with the finding by Bornmann and Daniel (2008). The current review also shows that to better understand the relationship between citing and cited documents, a variety of features should be analyzed, primarily the citation context, the semantics and linguistic patterns in citations, citation locations within the citing document, and citation polarity (negative, neutral, positive).

## Article Highlights

Computational and machine learning techniques have facilitated citation context/content analyses of large datasets in comprehensive studies.

The automated recognition of citation functions has the potential to enhance the information retrieval capabilities of search engines.

Papers are cited for different scientific and non-scientific reasons. Only a small percentage of citations are influential (important).

Evaluative bibliometrics (citation analysis) would profit from considering insights from citation context/content analyses to facilitate more meaningful results.

## 1 Introduction

For several decades, citation counts have been used as a main science indicator to measure the scientific impact and performance of departments and research institutions, universities, books, journals, nations (Bornmann & Daniel, 2008; Safer & Tang, 2009), as well as individual researchers for "hiring, promotion, and awarding grants and prizes" (Safer & Tang, 2009, p. 51). Citations can be used to present a historical overview of research areas as well as to project their future (Judge, Cable, Colbert, & Rynes, 2007). Citations play a significant role in understanding the link between scientific works that are somehow related to each other in terms of theory, methodology or result (Di Marco, Kroon, & Mercer, 2006). Citations have also been used in a few studies to examine the creative potential (novelty) of papers (Tahamtan & Bornmann, 2018b).

Citation analysis involves measuring the number of citations that a particular work has received, as an indicator of the overall quality of that work (Anderson, 2006). Citation analysis can also be used to recognize the areas worth funding (Safer & Tang, 2009). However, purely quantitative citation analysis has been widely criticized by researchers, arguing that citations should not be treated equally (Zhang, Ding, & Milojević, 2013). In the traditional citation analysis, citations are treated equally, while in practice they are based on different reasons and have different functions (Jha, Jbara, Qazvinian, & Radev, 2017; Zhang et al., 2013). For example, some cited papers are extensively discussed and others are arbitrarily or perfunctorily cited (Teufel, Siddharthan, & Tidhar, 2006). Giving all citations equal value overlooks the numerous potential functions they have for citing authors (Zhu, Turney, Lemire, & Vellino, 2015). Therefore, through conventional citation analysis, we are unable to identify the specific contribution of a given work to the citing work (Anderson, 2006).

Jha et al. (2017) noted that a more robust measure of citations is to use the citation context to provide additional information about how a cited paper has been used in the citing paper (Hernández-Alvarez, Gomez Soriano, & Martínez-Barco, 2017). In other words, to understand citation impact, an extended form of citation analysis has been used by researchers, which is known as citation content/context analyses (Hernández -Alvarez & Gomez, 2015). Citation content/context analyses have been proposed as complementary methods to traditional citation analysis (Zhang et al., 2013). Content/context analyses are "motivated by the need for more accurate bibliometric measures that evaluates the impact of research both qualitatively and quantitatively" (Abu-Jbara, Ezra, & Radev, 2013, p. 604). These methods have been used to produce a variety of citation function classification schemes. The schemes provide additional knowledge about the nature of the relationships between scientific works (Di Marco et al., 2006).

Analyzing the context of citations can be used to determine the extent and nature of the influence of a work on subsequent publications (Anderson, 2006). Citation context has been operationalized in several ways, including the position of the citation within the citing text, the semantics surrounding the reference (Bertin, Atanassova, Sugimoto, & Lariviere, 2016), and the words around citations (Bornmann, Haunschild, & Hug, 2018). Citation content analysis has also been used by some studies to determine the functions of citations. Here, the semantic content of the text surrounding the citation within the citing document is analyzed to characterize the cited work. One advantage of citation content analysis over pure citation analysis is that the former takes into account both the quantitative and qualitative factors (e.g. how one cites). Conventional citation analysis is quantitative in nature and does not consider actual content or context information (Zhang et al., 2013).

Over ten years ago, Bornmann and Daniel (2008) presented an overview of studies on citation content/context analyses, as well as the citing behavior of scientists. The study by Bornmann and Daniel (2008) covered the studies published from the early 1960s up to mid-2005. They attempted to address a core question: "What do citation counts measure?" They aimed to identify "the extent to which scientists are motivated to cite a publication not only to acknowledge intellectual and cognitive influences of scientific peers, but also for other, possibly non-scientific, reasons" (Bornmann & Daniel, 2008, p. 45).

Since then, technical developments have brought extensive changes to data availability and analysis over recent years. Reading a huge number of publications for context or content analyses purposes is a tedious task which requires dedicating a large amount of time and energy (McCain & Turner, 1989). However, technical developments have influenced the methods and techniques used in analyzing the contexts of citation. For example, sentiment analyses of citations via machine learning and other computational techniques have received a great deal of attention in recent years for categorizing citations (see, e.g. Teufel et al., 2006). Having access to full text databases has enabled researchers employing computational techniques to conduct complex analyses on scientific documents (Bertin, Atanassova, Sugimoto, et al., 2016).

The present review aims to update the review of Bornmann and Daniel (2008) with an additional focus on the technical developments in the last decade, which have facilitated studies of citations. For example, access to the machine-readable formats of scientific papers and automated data processing has provided bibliometric researchers with the opportunity to work with larger datasets, conduct large-scale studies, and employ new approaches and methods for studying citations (Bertin, Atanassova, Gingras, & Larivière, 2016).

## 1.1 Theoretical approaches to explaining citing behavior

In this section, we do not aim to present a comprehensive overview of the theories of citing behavior since these have already been explained in previous studies (see, e.g. Bornmann & Daniel, 2008; Nicolaisen, 2007; Tahamtan & Bornmann, 2018a). However, we will briefly explain these theories, together with several recent attempts to propose citation theories and models. These theories and models form the basis for citation context/content analyses and surveys on citing behavior. The two traditional theories are the normative and social-constructivist theories. The normative theory was proposed by Merton (1973), who explained that scientists primarily cite their peers to give them credit. According to normative theory, reasons to cite are of cognitive nature. The social-constructivist theory claims instead that peer recognition is not the only reason for citing. According to the social-constructivist theory, the citation decision process is multidimensional and depends on many factors. For example, the social-constructivist theorists believe that scholars cite scientific works to persuade readers that the claims they have made in their own scientific works are robust and valid (Nicolaisen, 2007). As such, scientists cite "to defend their claims against attack, advance their interests, convince others, and gain a dominant position in their scientific community" (Bornmann & Daniel, 2008, p. 49).

The normative and social-constructivist theories of citing have been widely critiqued. Some researchers have attempted to propose alternative citation theories or models overcoming the weaknesses of these two traditional theories. Nicolaisen (2007) is among such scholars who introduced a theory which has its roots in the handicap principle (proposed by Zahavi & Zahavi, 1999). According to Nicolaisen (2007), authors avoid careless and dishonest referencing because they are afraid of being criticized by their peers. As such, scientists try their best to honestly cite documents "to save the scientific communication system from collapsing" (Nicolaisen, 2007, p. 629).

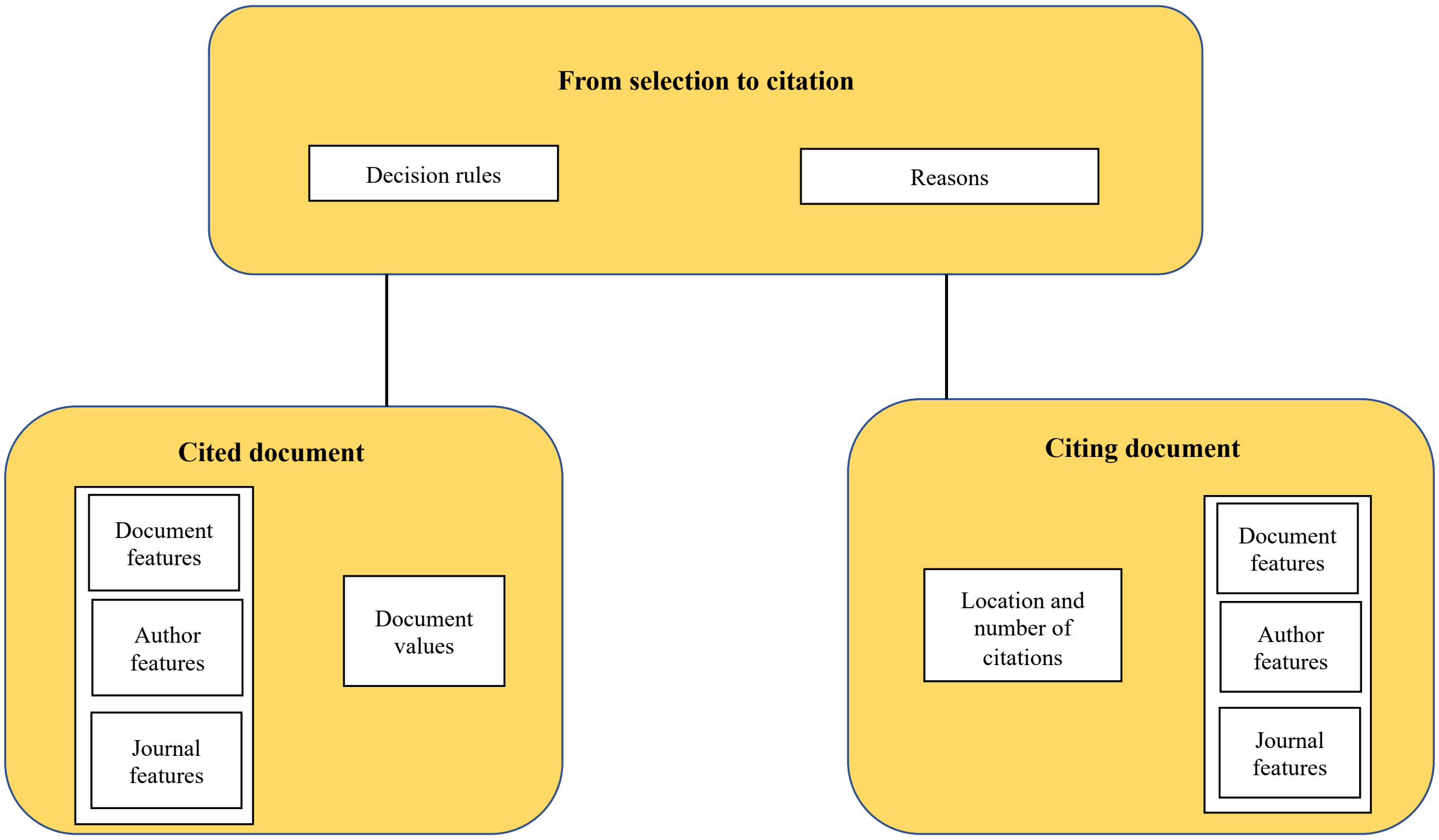

Figure 1. Core elements in the process of citing

Source: Tahamtan and Bornmann (2018a, p. 205)

To overcome the very diverging positions in previous citation theories, Tahamtan and Bornmann (2018a) proposed a synoptic model explaining the core elements in the process of citing. The model summaries previously published empirical studies on citing behavior. The model consists of three core elements: cited document, from selection to citation, and citing document (see Fig. 1). According to this model, selecting and citing a document is influenced by many factors, some of which are not subject to the control of the citing authors (e.g. the journal's or reviewers' requirements for citing certain documents). According to this model, documents are chosen to be cited in the citing document through a citation decision process. "This process is characterized by specific reasons to cite and decision rules of selecting documents for citing" (Tahamtan & Bornmann, 2018a, p. 205). Scholars' citing decisions are influenced by (many) factors that are related to both the citing and cited document. One major advantage of this conceptual model over previous citation theories and models is that (a) it is based on a comprehensive overview of empirical studies on citing behavior (it is a conceptual overview of the literature), and (b) it includes many of the identified reasons for citing from both the normative and social-constructivist camps.

## 1.2 Technical developments and new sources of citation studies

In the past, one main challenge in citation context studies was the great effort and time required to manually analyze and categorize the text around citations. Even when computational techniques were used to analyze the data, data processing of the PDF format of papers was problematic, tedious and time consuming (Bertin, Atanassova, Gingras, et al., 2016; Pride & Knoth, 2017). As a consequence, most citation content/context studies were carried out on small datasets (Bornmann et al., 2018).

However, nowadays, as a result of technical developments, such as the existence of machine-readable formats of publications (XML tags), recognizing the features of citation contexts have become much easier and faster (Bornmann et al., 2018; Hu, Chen, & Liu, 2015). The machine-readable formats of papers contain information about the exact locations of citations and the context in which the citations appear (Boyack, van Eck, Colavizza, & Waltman, 2018). The XML tags contain a variety of metadata information such as paper's title, authors, abstract, bibliography, and in-text citations. This means that each paper's content is now available in structured full text format, which makes automated text processing much easier than in the past (Bertin, Atanassova, Gingras, et al., 2016).

Using the full text of papers in machine-readable format has allowed researchers to study the different features of citations, such as the citation purposes and functions, citation polarity (negative, neutral, positive), citation locations (Boyack et al., 2018; Jha et al., 2017; Teufel et al., 2006), and the linguistic patterns in citation contexts (e.g. the distribution of words, verbs, and hedges) (Di Marco et al., 2006). As such, some studies have made use of the XML-formatted full text of papers to design citation function and/or citation polarity classifiers (e.g. Jha et al., 2017; Teufel et al., 2006).

Large-scale studies have also been made possible as a result of these technical developments. For example, Boyack et al. (2018) investigated the in-text citation distribution of over five million papers from two large databases – the PubMed Central and Elsevier journals. In most citation context studies, the citation locations are analyzed to provide a better understanding of the purposes for which references have been cited. The section structure of papers, IMRaD (Introduction, I, Methods, M, Results, R, and Discussion, D), is an important feature to be used in citation classifiers to improve their performance in detecting citation functions Bertin and Atanassova (2014).

Over recent years, many journals and publishers have made scientific papers available and downloadable in XML-formatted full texts (Bornmann et al., 2018; Boyack et al., 2018; Hu et al., 2015; Small, Tseng, & Patek, 2017). Elsevier, Springer, John Wiley & Sons, PLOS, PubMed Central, and Microsoft Academic are among the publishers/databases that provide XML-formatted full texts (Bornmann et al., 2018; Hu et al., 2015; Small et al., 2017).

Elsevier's ConSyn (http://consyn.elsevier.com) has provided the XML format for papers since 2011. Citation instances (sentences in which citations appear) can easily be recognized and extracted via ConSyn, because they are marked with XML tags (Hu et al., 2015). PLOS journals are great sources of citation content and context research, since they cover all fields of science and social sciences. In PLOS, papers are available in XML format (Bertin, Atanassova, Sugimoto, et al., 2016).

The Association for Computational Linguistics Archives (ACL) Anthology (https://www.aclweb.org/anthology/) has been used by many researchers to conduct citation content/context studies (e.g. Hassan, Safder, Akram, & Kamiran, 2018; Hernández-Alvarez et al., 2017; Jha et al., 2017; Valenzuela, Ha, & Etzioni, 2015; Zhu et al., 2015). CiteSeer (http://csxstatic.ist.psu.edu/) which contains publications in computer and information sciences, is another source that can be considered for citation context studies (Doslu & Bingol, 2016). Microsoft Academic is another valuable source of citation data for both papers and books (Kousha & Thelwall, 2018). It is a potential database for conducting citation context studies, because it has made it possible to download citation contexts that are already segmented (Bornmann et al., 2018).

## 2 Methods: search for the literature

To find the relevant literature on citation content/context analyses, and the surveys or interview studies on citation motivation, we used the methods explained in Tahamtan, Afshar, and Ahamdzadeh (2016) and Tahamtan and Bornmann (2018a). The search for the literature was conducted in 2019 and included the original English language papers in the period of 2006 to 2018. The publications of all document types were searched in WoS and Scopus using the following search strategy: *"citation classification" OR "citation context" OR ("content analysis" AND citation) OR "citation function" OR "in-text citation" OR "citation behavior" OR "citation behaviour" OR "citation motivation" OR "citer motives" OR "citing motives".* We limited our search to the title of documents in both databases to receive the most relevant documents. Our search strategy retrieved 188 papers: 124 from Scopus, 55 from WoS, and 9 from PubMed.

We imported the retrieved papers into Endnote and removed duplicate studies (n=57). The remaining 131 papers were screened by titles, abstracts, and full texts to exclude irrelevant or less-relevant papers. Overall, from our search in the three databases, 29 relevant documents were included and 102 irrelevant or less-relevant were excluded from the study. We found a few other relevant papers by browsing the bibliography of relevant papers. This resulted in a further 12 relevant papers for our review. Overall, 41 studies were included in the current review. When necessary, the authors of this study discussed the relevance of papers and whether or not they should be included in the review or not.

## 3 Empirical results of studies on citations

In the field of bibliometrics, analyzing and classifying citations has become an emerging research topic in recent years in order to understand authors' motivations for citing literature (Bakhti, Niu, Yousif, & Nyamawe, 2018) and to gain a better understanding of the relationship between citing and cited works (Bornmann & Daniel, 2008). In terms of methodology, two approaches have been employed to determine the reasons for citing or the functions of citations (Bornmann & Daniel, 2008):

1) Citation content/context analyses; and
2) Surveys or interviews with scientists on their citing motives and behaviors.

In order to obtain a summary of the literature, some of the main features in the 38 studies which were included in the current review were extracted and inserted into Table 1. These features included "data source", "sample size", "data processing method", "study objective", and "main results". The papers were classified into three groups (following the main approaches in the studies, see above): (1) content and context analyses of citations to characterize the cited documents, (2) citer motivation surveys or interviews, and (3) reviews of previous studies. The studies on citation content and context analyses were divided into two groups: "automated data processing", and "manual data processing".

The results of the studies in Table 1 are explained in detail in the following sections. The studies in the table are sorted by type of study (first citation context/content studies and second citer motivation studies), and – within the types – by publication year.

Table 1. Summary of the literature on citation content/context analyses, citer motivation surveys or interviews, and reviews

| Paper | Data source | Sample size | Data processing method | Study objective | Main results |
|---|---|---|---|---|---|
| Citation context/content study | | | | | |
| Anderson (2006) | Social Science Citation Index | 328 papers citing Karl Weick (578 citation contexts) | Manual | To identify the influence of Karl Weick's book on citing documents | The most frequently cited concept was "enactment" (16.6%).<br>Regional differences existed in scholars' citing behaviors. |
| Teufel et al. (2006) | Computation and Language E-Print Archive | 116 citing papers (and 2829 citation instances of these papers) | Automated | To identify citation functions/polarity | The annotation scheme achieved a degree of accuracy of at least 75% in determining citation functions and 83% in determining citation polarity. |
| Di Marco et al. (2006) | BioMed Central | 985 papers | Automated | To identify the distribution of hedge cues in citation contexts and across different paper sections | Hedge cues were more frequently observed in the citation contexts than the remaining text. |
| Siontis, Tatsioni, Katritsis, and Ioannidis (2009) | Web of Science | 15 citing papers | Manual | To identify the weaknesses of two clinical trials mentioned in the papers citing them | More than half of the cancer news stories had used an optimistic tone toward clinical trials, followed by neutral tone (40%), and pessimistic tone (9.8%). |
| Anderson and Sun (2010) | Social Science Citation Index | 301 papers citing Walsh and Ungson paper (496 citation contexts) | Manual | To identify the influence of the Walsh and Ungson paper on citing documents | The most disciplines citing this work were "management" (55%), and "information technology" (27%).<br>Only 3.4% of citation contexts were "critical". |
| Wang, Villavicencio, and Watanabe (2012) | IEEE transactions | 40 citing papers (345 citation contexts) | Automated | To identify citation functions | More than 50% of citation contexts were "extend", followed by "criticize" (30.14%), "compare" (13.88%), and "improve" (3.83%). |
| Danell (2012) | Web of Science | 178 citing papers | Manual | To identify the influence of three highly-cited papers in complementary and alternative medicine (CAM) on citing documents | 25% of the citing documents were classified as "medicine, general, internal".<br>The "positive/confirmatory" and "negative/critical" citations were relatively short and brief, without going into details. However, mixed citation contexts (e.g. positive/confirmatory + neutral/empty) were more detailed. |
| Ramos, Melo, and Albuquerque (2012) | Scopus | 212 citing papers (476 citation contexts) | Manual | To identify the influence of Phillips and Gentry (1993) and Bennett and Prance (2000) on citing documents | Most citing papers had minor relevance.<br>Citing authors barely read the documents or read them carelessly. |
| Li, He, Meyers, and Grishman (2013) | PubMed | 91 citing papers (6355 citation instances) | Automated | To identify citation functions/polarity | The classification scheme achieved a degree of accuracy of 67% in determining citation functions.<br>Among the functions "discover$^{+}$" had the highest precision (80%). |
| Halevi and Moed (2013) | ScienceDirect | 32 citing papers (1150 citation instances) | Automated | To determine how in-discipline and out-discipline citations have been distributed and used within different article sections | In-disciplinary citations were found with a higher frequency in the “methodology” section than other sections. |
| Chang (2013) | Web of Science | 908 papers citing de Solla Price (1142 citation instances) | Manual | To identify the cited concepts and citation functions between natural sciences (NS) and social sciences and humanities (SSH) | The top-cited concepts in both natural sciences (NS) and social sciences and humanities (SSH) were “science growth patterns”, “scientific communication”, and “scientific productivity”.<br>The top-cited functions in both natural sciences (NS) and social sciences and humanities (SSH) were “evidence”, “related studies”, and “background information”. |
| Åström (2014) | Web of Science | 6853 papers citing Gérard Genette's books and 234 papers citing “paratext” | Automated | To identify the influence of the Gérard Genette’s books on citing documents | The focus of papers citing “paratext” was mainly on “empirical concepts”, while the Gérard Genette’s books contained both “empirical” and “theoretical concepts”. |
| Sieweke (2014) | Social Science Citation Index | 352 papers citing Pierre Bourdieu (476 citation contexts) | Manual | To identify the influence of Pierre Bourdieu on citing documents | 46.6% of the citations of Bourdieu’s work came from three concepts: “capital”, “habitus”, and “field”. |
| Bertin and Atanassova (2014) | PLOS journals | 9446 citing papers (459834 citation instances) | Automated | To identify verbs and their distribution in citation contexts | The frequency of verbs in citation contexts vary in different sections of papers. |
| Galgani, Compton, and | Australasian Legal Information Institute | 2027 legal documents (3954 citations) | Automated | To identify citation functions | The level of human experts’ disagreement on labeling legal citations was high. |
| Hoffmann (2015) | | | | | The classification system performed better when the citations on which experts didn't agree were removed. |
| Valenzuela et al. (2015) | ACL Anthology | 20527 citing papers (106509 citations):<br>a sample of 465 cited – citing paper pairs | Automated | To identify influential (important) citations | 14.6% of the citations were labeled as important (i.e. had used or extended the cited work). |
| Zhu et al. (2015) | ACL Anthology | 100 citing papers (3143 papers cited) | Automated | To recognize influential references | The most predictive feature to identify influential papers was "the number of times a paper was cited in the citing paper".<br>10.3% of the 3134 cited references were influential and 89.7% were non-influential. |
| Liu, Ding, Wang, Tang, and Qu (2015) | Web of Science | 200 papers citing John O'Keefe's work (228 citation sentences) | Manual | To identify the influence of John O'Keefe's work on citing documents | The most frequent terms in citing sentences were "cell placement", "hippocampus", and "environment".<br>Most citations were neutral (n=204). |
| Zavrsnik, Kokol, Del Torso, and Blazun Vosner (2016) | Web of Science | 926 citing papers | Automatic | To identify the influence of five pediatric sleeping beauties on the papers citing them | The main contents in the citing papers were associated with the content discussed in each cited sleeping beauty. |
| McCain and | ISI files (dialog files) | 497 papers citing<br>Frederick P. Brook, Jr's | Manual | To identify the influence of Frederick P. Brooks, Jr's book on citing documents | The most cited concepts were "generalia" (general description of the book), "project management issues", and "building the system" (among others). |
| Salvucci (2016) | | (574 citation contexts) | | | |
| Bertin, Atanassova, Sugimoto, et al. (2016) | PLOS journals | 75000 citing papers (3 million citation sentences) | Automated | To identify the distribution of linguistic patterns in citation contexts in different paper sections | The most frequent words in the citation contexts varied according to their location in the text. |
| Jha et al. (2017) | ACL Anthology | 30 citing papers (3500 citations) | Automated | To identify citation functions/polarity | "Use" had the highest frequency (14.7%), followed by "criticism", "substantiation", and "basis". |
| Pride and Knoth (2017) | ACL Anthology | 415 citation pairs | Automated | To identify influential citations | The total number of times a paper was cited in single citing papers was a strong indicator of citation influence on citing papers. |
| Hernández-Alvarez et al. (2017) | ACL Anthology | 86 citing papers (2120 citation instances) | Automated | To identify citation functions/polarity | "Based on" was the most frequent positive function (n=255), "useful" was the most frequent neutral function (n=429), and "weakness" was the most frequent negative function (n=128). |
| Small et al. (2017) | PubMed Central | 128 citing papers (6574 citances) | Automated | To develop a classifier based on discovery words for identifying discoveries | Only 46% of the papers that had discovery words in their citances were scientific discoveries.<br>The classifier reached a high accuracy in recognizing discoveries (94%). |
| Lin (2018) | Taiwan Humanities Citation Index-Core, and Taiwan Social Sciences Citation Index | 360 citing papers (25617 citations) | Manual | To study how "essential" versus "perfunctory" and "confirmative" versus "negational" citations were distributed across six sub- | 97.13% of citations were "confirmative" and 2.83% were "negational".<br>Humanities had more "negational" citations than the social sciences. "Perfunctory" citations in these |
| | | | | disciplines in the humanities and social sciences (H&SS) | six H&SS subjects accounted for less than 50% of citations. |
| Cristea and Naudet (2018) | Scopus | 120 citing papers | Manual | To identify how Leucht, Hierl, Kissling, Dold, and Davis (2012) was cited in the literature | 60% of the citing papers had pointed to the argument that no substantial difference exists in the treatment effectiveness of psychiatry and general medicine. |
| González-Teruel and Abad-García (2018) | Web of Science and Scopus | 332 papers citing Elfreda Chatman | Manual | To identify the influence of Elfreda Chatman's theories on citing documents | Most of the citing works were in "social sciences", "computer science", and "medicine". |
| Small (2018) | PubMed Central | 1000 most cited papers in PubMed Central (646347 citances in these papers) | Automated | To investigate which citation context features (e.g. hedging words) best predicted whether a paper was method or non-method paper | The frequency of hedging words (may, show, not, and suggest) was higher in non-methods than methods papers. |
| Hassan et al. (2018) | PLOS ONE | 4138 papers (21804 cited references) | Automated | To identify important citations | Important citations were rare. For example, Austria with only 13.71% was the top country in terms of the total number of important citations, followed by Sweden (9.77%) and Brazil (9.11%). |
| Bornmann et al. (2018) | Microsoft Academic | 59 papers by Eugene Garfield (428 citation contexts) | Automated | To understand how Eugene Garfield's works have been perceived by citing documents | "Journal Impact Factor" was the most frequent keyword in three networks.<br>Eugene Garfield's papers have been cited in the context of a wide range of topics (e.g. bibliometric data, indicators, units of bibliometric analysis, bibliometric methods). |
| Bakhti, Niu, and Nyamawe (2018) | ACL Anthology | 300 papers (8700 citation sentences) | Automated | To identify citation functions | The most frequent function was “useful” (24.81%).<br>Considering authors’ information improved the performance of the classifier in identifying citation functions. |
| Citer motivation study | | | | | |
| Clarke and Oppenheim (2006) | - | 65 postgraduate students | - | To study the citing behavior of postgraduate students | Students citing decisions were more frequently based on the relevance and importance of the cited works than other (non-scientific) reasons. |
| Tang and Safer (2008) | - | 50 psychologists and 49 biologists | - | To identify the importance of cited references, and textual features of cited references to predict citation importance | Citations were rated “fairly important”.<br>The most frequent reasons for citing were “general background” (37.3%), “conceptual ideas” (31.0%), and “methods and data” (13.4%). |
| Harwood (2008a) | - | Six computer scientists and six sociologists | - | To identify the reasons for which names of cited authors were mentioned in different formats | The “publisher policy”, “editor’s decision”, “co-authors”, and many other factors influenced the format with which the names of cited authors were mentioned in the text. |
| Harwood (2008b) | - | Six computer scientists and six sociologists | - | To identify the impact of the publication outlets on authors’ citation patterns | The outlet in which the publication of the author was published influenced their citing behaviors.<br>The “outlet "policy”, “co-authors”, “outlet type”, “outlet’s favored research paradigm”, etc. influenced authors’ citing behavior. |
| Harwood (2009) | - | Six computer scientists and six sociologists | - | To identify the citer motivation in citing the literature | The most frequent function was “position” in sociology (46.2%) and “signposting” in computer science (26%). |
| | | | | | Some citations had more than one function. |
| Milard (2014) | - | 32 French chemists | - | To identify social relations of citing authors with cited authors | Purely social references were rare (20 out of 1410 references). |
| Thornley et al. (2015) | - | 87 researchers | - | To identify the reasons for which cited authors (references) were regarded as trustworthy | The main reason for citing was “knowing the cited authors” (24.16%).<br>In 15.1% of the cases, papers were cited because they were “classical” or “seminal works”. |
| | | | Review | | |
| Camacho-Miñano and Núñez-Nickel (2009) | Several databases (the names of the databases are not presented) | 139 papers | - | To review the studies addressing why authors select some papers in preference to others. | Presented a model for citation decisions including three phases, (1) external limitations, (2) functional choice and (3) preferential selection. |
| Erikson and Erlandson (2014) | - | - | - | To propose a taxonomy of motives to cite. | Proposed a taxonomy of motives to cite with four categories including “argumentation”, “social alignment”, “mercantile alignment”, and “data”. |

### 3.1 Content and context analyses of citations to characterize the cited works

Citation content and citation context studies are based on an analysis of the text (within a sentence) around a reference anchor. Some studies have analyzed at least one sentence before and after the sentence including the citation, since the discourse regarding the cited paper often either continues beyond the citing sentence or a few sentences before it (Jha et al., 2017). Halevi and Moed (2013) have recommended an analysis of the sentence in which the citation appears, as well as one sentence before and after the sentence including the citation. Abu-Jbara et al. (2013) showed that in the 3500 citation contexts they studied, 22% consisted of two or more sentences. Some studies have attempted to identify the optimal size of context windows (for a review see Hernández -Alvarez & Gomez, 2015). For example, Ritchie, Robertson, and Teufel (2008) compared different citation context sizes and found that longer citation contexts performed better than short contexts (three sentences compared to one sentence) for information retrieval purposes. Similarly, Yousif, Niu, and Nyamawe (2018) showed that a larger citation context size (four sentences) increased the classification performance (in detecting citation purposes and sentiments) more than smaller citation context sizes (one sentence).

The differences between citation content and citation context studies are sometimes difficult to identify. Figure 2 illustrates and differentiates between citation content analysis and citation context analysis.[1] In **citation content analysis**, the semantic content of the text surrounding a given citation (cited document) within the citing document(s) is read to characterize the cited document (Halevi & Moed, 2013; Harwood, 2009; Liu, 1993; McCain & Turner, 1989). For example, in citation content analysis, the reasons for the influence of a certain author's publications can be determined. In other words, it can be determined for what reasons the documents have been cited. Zhang et al. (2013) noted that citation content analysis can be used to describe the contextual relationship between citing and cited documents, to examine the social and intellectual interactions between the citing and cited authors, and to understand the functions of citations. The studies that have used citation content analysis are mentioned in section 3.1.2.1. These studies have mainly analyzed the citation content manually. For example, Anderson (2006) investigated the influence of Karl Weick's book on citing documents. 578 citations (sentences in which Karl Weick's book appeared) were analyzed and categorized to 12 frequently cited concepts. It was found that the most frequently cited concept was "enactment" (16.6%). This study further demonstrated that Karl Weick was cited differently (for different reasons) by US-based journals versus European-based journals.

In **citation context studies**, the citing text around the reference anchor is analyzed. In other words, the text around citations (cited documents) in the citing document is used to characterize citations in the citing document. It is not the objective of citation context studies to yield information about the content of a certain cited document, but to characterize the citation process of the citing authors. The text is used to analyze the amount of multiple citations of the same document in the citing publication, to determine the different functions citations may have (e.g. giving credit or rhetorical devices), and to investigate the different meanings of citations (e.g. perfunctory or confirming). Many studies of this type take into account the location of citations within the citing document (e.g. in specific sections) to ascertain citation functions (Halevi & Moed, 2013). These studies often use automated data processing to analyze citation contexts.

[1] The figure also includes citer motivation surveys or interviews, which are explained in section 3.2.

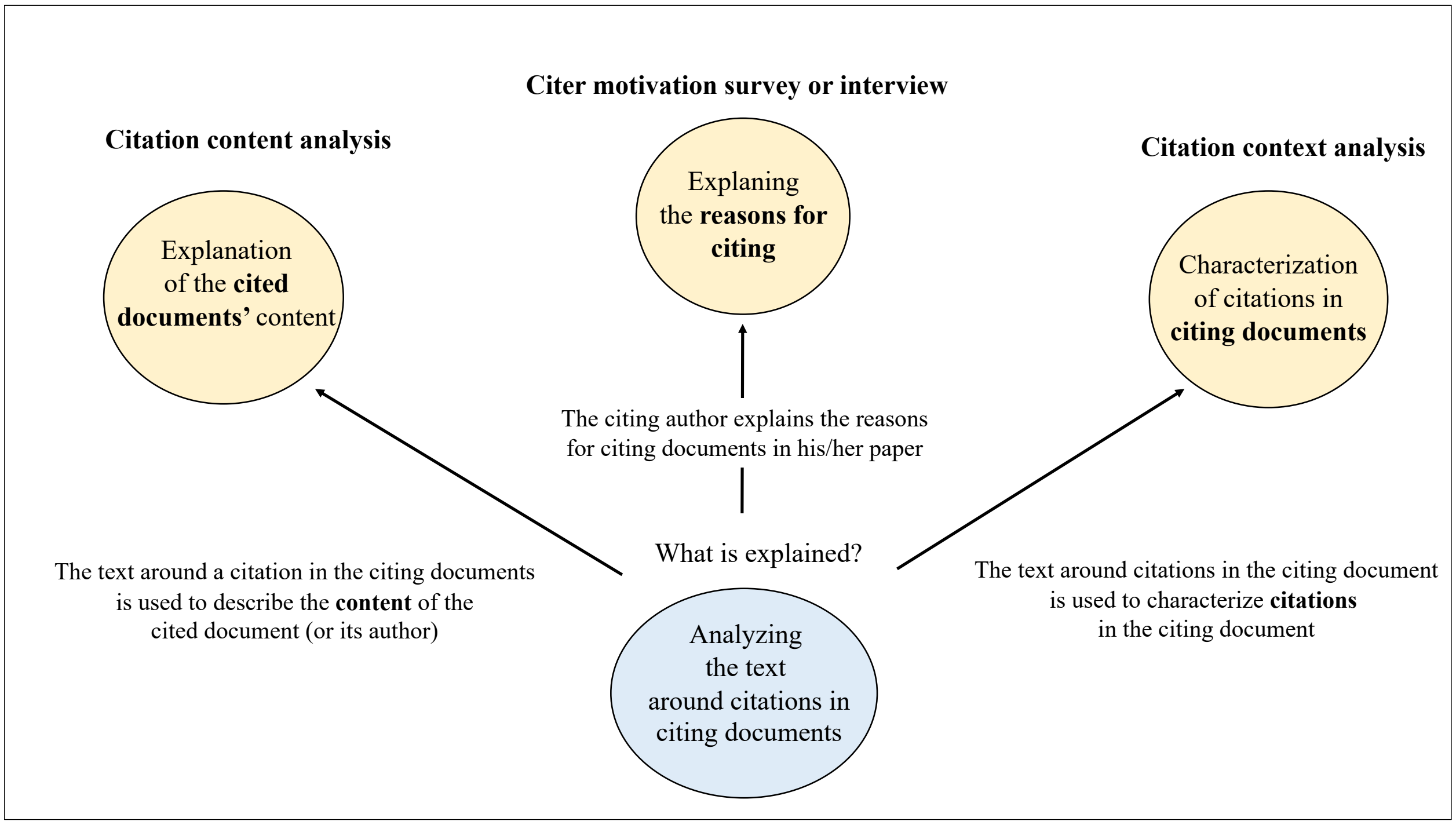

Figure 2. Definitions and components of citation content analysis, citation context analysis, and citer motivation survey/interview.

Citation content/context analyses have their own challenges, some of which have briefly been explained by Halevi and Moed (2013). For example, one major challenge of citation content/context analyses is that contradictory results are obtained, mainly because they are often achieved manually based on the subjective judgment of scientometricians who in many cases are not experts in the area under study (Halevi & Moed, 2013). Another issue here is that a sentence may contain several citations, while "parts of a sentence may not be talking about a cited paper even if it contains a reference anchor to it" (Jha et al., 2017, p. 3). Another issue is that even a citation that appears once in a paper can have more than one function (Erikson & Erlandson, 2014).

In the following section, citation content/context studies are presented which were published in recent years (2006 to 2018). The studies are categorized into two groups, according to their data processing and analysis methods: "automated data processing" and "manual data processing". The automated data processing category includes the papers that have partially used computational techniques and automated data processing methods to analyze the content/context of citations (annotated citation contexts in XML format). Studies of this type often use machine learning experiments to replicate the human annotation (Teufel et al., 2006). The manual data processing category includes the papers that have not used such computational methods for analyzing the content/context of citations. Our review indicated that many studies in the category of manual data processing had used citation content analysis to investigate the influence of highly-cited classical works on citing documents.

Both the automated and manual data processing methods have their own pros and cons. The main issues with manual annotation are that it is time consuming, tedious, and difficult. The issue with the studies that use automated data processing is that the classification schemes they propose for identifying citation functions do not yield reliable results (Hernández-Alvarez et al., 2017).

### 3.1.1 Automated data processing

We classified the automated data processing studies into four categories: (1) citation function and/or polarity (sentiment), (2) linguistic patterns (hedge cues, verbs, and words), (3) influential versus non-influential citations, and (4) other studies. Citation function and/or polarity refers to the studies that have attempted to create citation function/polarity classifiers. The second category includes the studies that have used hedge cues, verbs, and words to analyze citation contexts. The third category comprises the papers that have studied citation contexts to classify them as influential or non-influential. The studies that could not be classified in the previous categories are presented in the fourth category (other studies).

#### 3.1.1.1 Citation function and/or polarity

Teufel et al. (2006) were among the pioneer researchers who designed automated citation and polarity classifications. They studied citation functions and citation polarities independently, an approach that was later criticized by several studies (Jha et al., 2017; Li et al., 2013). Li et al. (2013) emphasized that in order to obtain a better understanding of the exact function of a citation, citation polarities should be incorporated and studied along with citation functions. Jha et al. (2017) supported this idea and noted that the relationship between cited and citing documents as well as the impact of scientific papers would be better recognized by determining both the citation polarities and citation purposes for the sentence in which a citation appears. These ideas were later followed by other researchers as well. For example, Hernández-Alvarez et al. (2017) mapped citation polarities to citation functions (see Table 4 below) and demonstrated the most frequent positive, neutral and negative functions. Hernández-Alvarez et al. (2017) defined citation influence in their classification based on citation polarity and citation function (see Table 5 below). In the following, we explain these studies to clarify their similarities and differences in terms of methodology, data sets, and classifications used.

Teufel et al. (2006) specified a classification scheme with four main categories: (1) weak: citations that had pointed to the weakness of previous studies, (2) contrast: comparisons or contrasts between the citing and cited works, (3) positive: positive sentiment about the cited work, and (4) neutral: citations that neutrally had described the cited work. The dataset used in the study by Teufel et al. (2006) consisted of 116 randomly selected conference papers and the citation instances (n=2829) in these papers from the Computation and Language E-Print Archive (http://xxx.lanl.gov/cmp-lg). The most frequent citations were “neutral citations” (more than 60%), and the least frequent citations were “negative citations” (4.1%). To automatically classify citation functions, several features, such as the cue phrases and citation locations were also taken into consideration. Their classification scheme achieved a degree of accuracy of at least 75% in identifying citation functions for all four categories (weak, contrast, positive, neutral). In other words, the proposed classifier could recognize citation functions with a degree of accuracy of 75%. Teufel et al. (2006) further conducted a citation polarity analysis with three categories (negative sentiments, positive sentiments, and neutral sentiments). Findings indicated that the classification scheme could determine the sentiments of citations with a degree of accuracy of 83%. Higher accuracy means greater success of the classification scheme in detecting citation functions and citation polarity.

A similar approach to Teufel et al. (2006) was used by Li et al. (2013). However, Li et al. (2013) took into account both citation purpose and citation polarity simultaneously for the same sentence. They proposed a citation classification scheme in which each function was labeled as either positive (+), neutral (=), or negative (-). The scheme included eight positive functions, three neutral functions and one negative function. Table 2 describes these categories. The dateset included 91 papers from PubMed along with their citation instances (n=6355). The overall distribution of "neutral" (citations that didn't carry information) and "negative" function (had a frequency less than 5) was 0.443. The sum of the distribution of all other citation functions was 0.557 in the dataset. "Co-citation$^{=}$" (0.333) and "Discover$^{+}$" (0.123) had the highest distribution. The proposed citation classification scheme achieved a degree of accuracy of 67% in detecting citation functions. Among all the studied functions, "Discover$^{+}$" had the highest precision (80%). Li et al. (2013, p. 406) noted that the differences in the classifier performance in detecting citations could be due to "the imbalance distribution of citation functions in the annotated corpus".

Table 2: Annotation scheme for citation function: + represents positive sentiment, = represents neutral sentiment, and − represents negative sentiment

| Citation function | Description |
|---|---|
| Based on$^{+}$ | A work is based on the cited work |
| Corroboration$^{+}$ | Two works corroborate each other |
| Discover$^{+}$ | Acknowledge the invention of a technique |
| Positive$^{+}$ | The cited work is successful |
| Practical$^{+}$ | The cited work has a practical use |
| Significant$^{+}$ | The cited work is important |
| Standard$^{+}$ | The cited work is a standard |
| Supply$^{+}$ | Acknowledge the supplier of a material |
| Contrast= | Compares two works in a neutral way |
| Co-citation= | Citations that appear closely |
| Neutral= | The cited work not belonging to other functions |
| Negative- | The weakness of the cited work is discussed |

Source: Li et al. (2013, p. 403)

Similarly, Jha et al. (2017) developed an annotated dataset for recognizing the context of citations which contained both the citation purpose and citation polarity for the same set of sentences. The dataset in this study contained 30 papers and their citations (n= 3500) from ACL Anthology. A citation was marked positive if (a) it explicitly stated a strength of the cited paper, or (b) the cited paper was used by the author or any other researcher, or (c) if the cited work was compared to another paper and was assessed better in some way. A citation was marked negative if (a) it pointed to a weakness of the cited paper, or (b) it was compared to another paper and was considered weaker in some way. Neutral citations were those that were only descriptive. A taxonomy consisting of six categories (criticizing, comparison, use, substantiating, basis, and natural) was created. Jha et al. (2017) used several features to classify citation purposes and citation polarity, such as the number of references in a citation context, whether a reference appeared alone in the citation context or with other references, whether the citation context contained a nagation cue, and whether the reference was a self-citaion.

The most frequent citation purpose was "use" (17.7%). In other words, in 17.7% of cases, the citing papers had used the methods, ideas or tools of the cited papers. Other citation purposes were "criticism" (14.7%), "comparison" (8.5), "substantiation" (7%), and "basis" (5%). Other categories accounted for 47% of citations. The overall accuracy of the classification scheme in identifying citation purposes was 70.5%. Among all citation purposes, the accuracy of recognizing "use" (60%) was higher than that of other categories. 30% of citations were "positive", 12% were "negative", and 58% were "neutral". The overall accuracy of detecting citation polarity was 84.2% (84.2% for neutral, 69.8% for negative, and 55.4% for positive citations).

When citation purposes were mapped to citation polarity, Jha et al. (2017) found some correlations between purpose categories and polarity categories (see Table 3). For example, the "negative polarity" had a higher frequency in the "criticizing purpose" category, both of which are negative in nature. Or the "basis" and "use" categories, both of which are positive in nature, led to a "positive polarity". These results exemplify the importance of utilizing both citation polarity and citation function simultaneously which was already emphasized by Li et al. (2013).

The polarity classifier designed by Jha et al. (2017) achieved a degree of accuracy of 90.1% for identifying negative and positive citations. This study showed that taking into account purpose categories improved the performance of the classifier in more accurately detecting negative sentences.

Table 3. Distribution of the citations belonging to different citation purpose categories across polarity categories

| Purpose label | Neutral | Positive | Negative |
|---|---|---|---|
| Criticizing | 0% | 33% | 67% |
| Comparison | 67% | 17% | 15% |
| Use | 26% | 73% | 0% |
| Substantiating | 1% | 99% | 0% |
| Basis | 20% | 80% | 0% |
| Neutral | 98% | 1% | 0% |

Source: Jha et al. (2017, p. 98)

Hernández-Alvarez et al. (2017) proposed a comprehensive citation classification scheme with three features: (1) citation function, (2) citation polarity (negative, positive, and neutral), and (3) citation influence classification. The authors noted that taking into account these features could create a more accurate index to measure the influence of cited papers on citing papers. Hernández-Alvarez et al. (2017) obtained citation instances (n=2120) of 86 papers from ACL Anthology and classified them in terms of their influence.

The preliminary citation functions were classified as "use", "comparison", "critique" and "background". However, several aspects were added to each citation function to attain more precise functions. For example, the aspects of "comparison" were: "comparison results are positive", "comparison results are negative", and "comparison results are neutral". After applying the aspects, citations functions were as follows: "based on", "supply", "useful", "acknowledge, "corroboration", "contrast", "weakness", and "hedges".

The authors mapped polarity to functions (see Table 4) and found that "based on" was the most frequent positive function (n=255), "useful" was the most frequent neutral function (n=429), and "weakness" was the most frequent negative function (n=128).

Table 4. Polarity mapped to function

| Function | Neutral | Positive | Negative |
|---|---|---|---|
| Acknowledge | 47 | 719 | 15 |
| Corroborate | 19 | 4 | 0 |
| Useful | 219 | 492 | 0 |
| Contrast | 13 | 72 | 21 |
| Weakness | 0 | 0 | 128 |
| Based on | 255 | 35 | 0 |
| Supply | 26 | 26 | 0 |
| Hedge | 0 | 0 | 38 |

Source: Hernández-Alvarez et al. (2017, p. 578)

The frequency of "hedge" was zero in both positive and neutral polarity, but not in negative polarity (n=38). Hernández-Alvarez et al. (2017). This is in line with the findings of Mercer, Di Marco, and Kroon (2004) who mentioned that hedge cues "may help to determine the purpose of citations, especially if they have a negative sentiment" (Hernández-Alvarez et al., 2017, p. 568).

In the study by Hernández-Alvarez et al. (2017), influence classification had three categories (see Table 5). The results indicated that the frequency of "perfunctory" citations was greater than "significant positive" and "significant negative". The proposed classification scheme received high accuracy for citation functions (89%), citation polarity (93%), and citation influence (94%) (Hernández-Alvarez et al., 2017).

Table 5. Proposed influence classification scheme

| Influence category | Description |
|---|---|
| Perfunctory | Citation is trivial, only marginally related to citing paper, often related to neutral polarity |
| Significant positive | Relevant citation, influential paper with positive polarity |
| Significant negative | Relevant citation, influential paper with negative polarity |

Source: Hernández-Alvarez et al. (2017, p. 571)

A closer look at the citation classifications indicate that some citation functions are common among different studies. For example, "use" can be seen in several studies (Bakhti, Niu, & Nyamawe, 2018; Hernández-Alvarez et al., 2017; Jha et al., 2017) which happens to have a high frequency compared to other functions. "Use" has also been found to be among the most frequent verbs in different articles sections types (Bertin & Atanassova, 2014). In this regard, Small (2018) showed that "using" had a high frequency in method papers and a low frequency in non-method papers.

#### 3.1.1.2 Linguistic patterns (hedge cues, verbs and words)

Other elements that could be used to determine the relationship between citing and cited documents are linguistic patterns (features). These linguistic patters were hedge cues (e.g. would, kind of, likely, kind of, mostly, literally, and although), verbs, and words. The linguistic patters of citations may differ according to their functions. "For example, citation sentences describing background of work are usually in active voice, while basic methods or tools used in the papers are in most cases introduced in passive voice" (Dong & Schäfer, 2011, p. 625). The linguistic patterns may also have varied distributions across different paper sections. Hedge cues have different occurrences in different paper sections (Di Marco et al., 2006). Some studies have investigated the usage of hedge cues in identifying the purpose of citations. One of the first studies that had explored hedge cues to determine citation purposes was conducted by Di Marco et al. (2006). Hedge cues are used within different sections of a paper for varied purposes. For example, they are used in the introduction section to refer to previous literature and highlight the significance of the work (Di Marco et al., 2006). (Di Marco et al., 2006) investigated how hedge cues were distributed across different paper sections (background, methods, results/discussion, and conclusions). They studied 985 biology papers from the BioMed Central. This study classified sentences into four categories: citation sentence, citation frame (the sentence next to the citation sentence), normal sentence, and hedge sentence (sentences containing hedge cues).

Results indicated that the frequency of all sentence types was higher in results/discussion, followed by methods, background, and conclusion. The frequency of hedge sentences was considerably higher in results/discussion than other sections. Overall, this study showed that hedge cues were extensively used in the sentences in which the citations appeared, as well as their surrounding sentences. It was also found that hedge cues were more frequently observable in the citation contexts than the whole text. Di Marco et al. (2006) noted that these findings indicate that hedge cues (along with other citation context features) are potential elements to be used in determining the pragmatic functions of citations.

Besides the importance of hedges in determining citation functions, a few studies have investigated the most frequently mentioned verbs in citation contexts to determine the relationship between citing and cited documents (Bertin &

Atanassova, 2014). It is suggested by Abdullatif, Koh, Dobbie, and Alam (2013, p. 30) that creating "a standard citation scheme requires accurate selection of verbs relevant to references in citation sentences". In another study, it is proposed that "citation contexts provide two kinds of vocabulary: technical words and words that signal sentiments or characterizations of prior knowledge" (Small, 2011, p. 376).

Wang et al. (2012) used 48 groups of cue phrases (based on verb, noun, and preposition) to detect and classify citation functions. The citation functions were "extend", "criticize", "improve", and "compare". 24 groups of cue phrases were used to find "extend"; 12 to find "criticize", 8 to find "compare", and 5 to find "improve" (see Table 6). The relationship between cited and citing documents were visualized in a citation network.

Table 6. Cue phrases in four citation types.

| Citation type | Cue phrases |
|---|---|
| Criticize | few of, little of, can only, however,<br>but, unfortunately, nevertheless, although, yet, nonetheless, limited to, restricted to |
| Extend | extend, use, rely, a descendent of, employ,<br>build on/upon, utilize, experiment on/with, combine, according to, apply, derived from, come from, benefit from, borrow, follow, adaptation of, obtained from, choose, inspired by, taken from, based on/upon, adopt, take it one step further |
| Improve | improvement, enhancement, better than<br>to avoid this problem, to solve this problem |
| Compare | different from, agreement with, compared to, like, similar to, in contrast to, unlike, identical to |

Source: Wang et al. (2012, p. 11)

In this study, 40 papers (345 citation contexts) from IEEE Transactions, published by Computer Society Digital Library, were analyzed. More than 50% of citation contexts were "extend", followed by "criticize" (30.14%), "compare" (13.88%), and "improve" (3.83%). Overall, the precision of the classification scheme in detecting citation functions was 62%. The precision of classification scheme in detecting "extend" (72%) and "criticize" (60%) was higher than "improve" (33%) and "compare' (29%).

This study "found that most writers have the same writing style when they criticize a research or show their intentions of utilizing the research" (Wang et al., 2012, p. 18). Wang et al. (2012) noted that increasing the number of cue phrases for each function in the classification scheme would lead to recognizing more correct relationships between cited and citing documents. The authors noted that, however, the appearance of multiple cue phrases in the same sentence may result in the low precision of the classification scheme in detecting citation functions (Wang et al., 2012).

Bertin and Atanassova (2014) presented an approach for identifying verbs in citation contexts, using a corpus of 9446 papers (459834 citation contexts) published by PLOS journals: *PLOS Biology*, *PLOS Computational Biology*, *PLOS Genetics*, *PLOS Neglected Tropical Diseases*, and *PLOS Pathogens*. Verbs in the citation contexts were identified and ranked according to the frequency of their occurrence in each paper section. Table 7 shows that certain verbs occurred many times within different sections of a paper. For example, Bertin and Atanassova (2014, p. 8) showed that in the introduction section, “70 verbs account for 50% of all verb occurrences, and 486 verbs account for 90% of the occurrences”. The word “show” was the most frequent verb in the introduction and discussion sections, and the second most frequent verb in results, but not among the top 10 verbs in the method section. “Show” was also among the top words in introduction and methods sections in the study by Bertin, Atanassova, Sugimoto, et al. (2016). It also was among the top words in discovery citation sentences (citances) of Small (2018) which are explained in the following. Other frequent words were “use”, “suggest, “include”, “perform” “follow”, “report”, and “obtain” (Bertin & Atanassova, 2014).

Bertin and Atanassova (2014, p. 11) noted that the density of some verbs in certain sections of a paper “confirms the hypothesis that citations play different roles according to their position in the rhetorical structure of scientific articles”. The most frequent verbs in each paper section are presented in Table 7 (Bertin & Atanassova, 2014).

Table 7. Top 10 of the most frequent verbs in the four section types

| Rank | Introduction | Method | Result | Discussion |
|---|---|---|---|---|
| 1 | show | use | use | show |
| 2 | use | perform | show | suggest |
| 3 | include | follow | find | use |
| 4 | suggest | obtain | report | report |
| 5 | identify | generate | observe | find |
| 6 | find | base | suggest | include |
| 7 | require | determine | identify | observe |
| 8 | associate | contain | express | require |
| 9 | involve | calculate | see | associate |
| 10 | lead | carry | include | involve |

Source: (Bertin & Atanassova, 2014, p. 4)

In another study, Bertin, Atanassova, Sugimoto, et al. (2016) proposed a natural language processing approach in order to recognize linguistic patterns in citation contexts. They assessed whether linguistic patterns varied according to citation locations. 75000 papers published by seven PLOS journals were analyzed. Similar to the previous study by two co-authors (Bertin & Atanassova, 2014, p. 4), Bertin, Atanassova, Sugimoto, et al. (2016) found that linguistic patterns in the citation contexts varied according to their locations. The verbs “show, “suggest, “find”, “know”, “demonstrate”, “include”, and “propose” were the top words in the introduction and methods sections. However, they did not occur frequently in the methods and results sections. The most frequent words in both the introduction and discussion were “observe” and “report”. The top words in the methods section were “describe”, “perform”, “calculate”, and “obtain”. The methods and introduction sections also had a high occurrence of “use” and “follow”. The distribution of “negations” (negative citations) was examined according to the frequency of occurrence of the negative word “not”. “Negation” was infrequent in the citation contexts. Similar to the distribution of “agree”, the frequency of “not” was higher in the results and discussion sections. Bertin, Atanassova, Sugimoto, et al. (2016) mentioned that their findings may have some implications for the construction of similarity indices for information retrieval purposes. They also pointed to one major limitation of this kind of citation context analysis: it does not distinguish between the nature of citations, namely citations which are perfunctory, confirming or other types of citations.

In two studies, Small et al. (2017) and Small (2018) used linguistic features in citances in a more practical way than previous studies (Bertin & Atanassova, 2014; Bertin, Atanassova, Gingras, et al., 2016; Di Marco et al., 2006). Previous studies had mainly attempted to determine the frequency and distribution of hedge cues, words, and verbs in different article sections. However, Small et al. (2017) and Small (2018) utilized a set of words in citances to classify papers to discovery and non-discovery papers, and method and non-method papers, respectively.

Small et al. (2017) investigated whether a set of words in citances that denote “discovery” could be used to identify biomedical discoveries. They defined a citance as a single sentence in which one or more references appear. They distinguished it from citation context definitions which may include various numbers of sentences before or after the citation. Small et al. (2017) analyzed a list of 128 biomedical discoveries (published in papers) retrieved from the PubMed Central and investigated whether the citances of those papers included terms such as “discovery”, “discover”, and “discovered”.

Discoveries were classified as either “violation”, “innovation”, or “extension”. A violation discovery was when an alternative view was proposed, in contrast to the accepted viewpoints in the scientific community. An innovation was when the study led to a new understanding of a phenomenon through unexpected findings. Extension referred to the studies that built upon prior discoveries. 16 violations, 71 innovation, and 41 extensions were found among the 128 papers. Only 46% of the papers that had discovery words in their citances were a scientific discovery. “Cause” “discovered”, “mechanism”, “first”, “important”, “recently”, “demonstrated”, “shown”, “reported”, and “found” appeared with a higher frequency in discovery citances than non-discovery citances. The most frequently mentioned non-discovery words were “algorithm”, “value”, “version”, “tool”, “analysis”, “using”, “data”, “performed”, “project”, and “used”. Small et al. (2017) created a classifier based on citance words. Results demonstrated that the classifier was able to recognize discoveries with a high accuracy (94%).

In another study, Small (2018) investigated which hedging words in citances best predicted method papers. The top 1000 most cited papers (646347 citances) indexed in PubMed Central were retrieved and classified as either a method or a non-method paper (55% of the 1000 papers were methods). This study demonstrated that the percentage of citances that contained hedging words, such as “may”, “show”, “not”, and “suggest” was higher in non-method papers than method papers. However, “using” along with some other non-language variables such as “age” (the publication year of the paper), “consensus”, and “section” had a higher frequency in method papers. Consensus was defined as the “mean cosine similarity of each citance for a paper with its cumulation of citances” and section was a “percentage of citances for a paper appearing in ‘method’ sections” (Small, 2018, p. 467). “Using” had a mean of 42.66 in method papers and a mean of 6.32 in non-method papers. Logistic regression was used to determine how much each variable predicted whether a paper was method or non-method. Results revealed that the predictive ability of “using”, with a degree of accuracy of 89.5%, was higher than other variables, followed by “may” (83.3%), “suggest” (76.6%), “show” (75.8%), "consensus" (71.5%), and "not" (68%). This study further investigated the accuracy of "word combination" (e.g. using and may), "consensus", and "section" variables in predicting whether a paper was method or non-method. The highest accuracy was obtained for a combination of "using + may + consensus + not + show + suggest", with a degree of accuracy of 92%. The second combination variables with the highest predictive power were "section + using + may + consensus + not + show", with a degree of accuracy of 91.9%.

#### 3.1.1.3 Influential versus non-influential citations

Another set of papers have employed a similar approach to Small et al. (2017) and Small (2018) in order to classify papers to two groups – influential (important) and non-influential papers (Hassan et al., 2018; Pride & Knoth, 2017; Valenzuela et al., 2015; Zhu et al., 2015). However, the main difference is that Small et al. (2017) and Small (2018) have mainly focused on linguistic cues to classify papers, while the studies that are described in the following have employed a wide range of features such as "the total number of times a paper was cited" in the citing paper in order to classify studies to influential and non-influential papers. The results of these studies show that most citations are non-influential and only a small proportion of them are influential. In the studies by Valenzuela et al. (2015), Zhu et al. (2015), and Pride and Knoth (2017) – presenting in the following – approximately only less than 15% of the citations were influential.

Valenzuela et al. (2015) introduced a classification approach for identifying influential (important) citations. Important citations were defined as the citations that had used or extended the cited work in a meaningful manner. Citations that appeared in the related work section or citations that were used to compare/contrast results were labeled as incidental citations. The dataset included 20527 papers and the 106509 citations in these papers from the ACL anthology. Valenzuela et al. (2015) annotated 465 cited-citing paper pairs and ordered them according to their importance. The citation classifier designed in this study found that only 14.6% (69 citation pairs) of the citations were influential and 85.4% (396 citation pairs) were incidental. They found that the "total number of times a paper was cited per section" (and in the entire text) in the citing document, and "author overlap" (self-citation) were the best predictors of academic influence on citing papers. Their classifier had a degree of precision of 65% in recognizing important citations. Valenzuela et al. (2015) noted that the classifier could be used in search engines to indicate which documents are important.

Zhu et al. (2015) used machine learning to identify the cited references that have been influential to the citing paper. The dataset included 100 papers and 3143 references cited in them. Similar to the study by Valenzuela et al. (2015), only a small proportion of citations were influential (10.3%), and 89.7% were non-influential. Results indicated that the most predictive feature to identify influential papers was "the total number of times a paper was cited in the citing paper", with a degree of accuracy of 35%. The accuracy of the classifier increased to 41% when this feature (the total number of times a paper was cited in the citing paper) was combined with "the number of different sections in which a reference appears". Adding the "self-citation" information to the model slightly improved the accuracy of the classifier (42%) in recognizing influential citations (see Table 3 in Zhu et al., 2015).

A similar study by Pride and Knoth (2017) used a dataset of 465 citation pairs from ACL Anthology to classify citations to influential and non-influential citations. 396 (85.7%) of references were found to be incidental and 69 (14.3%) were marked influential. Similar to the study by Valenzuela et al. (2015) and Zhu et al. (2015), the "total number of times a paper was cited in the citing paper" was a strong indicator of citation influence. This study showed that, contrary to the study by Valenzuela et al. (2015), "self-citation" and "similarity between abstracts" were other predictors of citation influence.

Hassan et al. (2018) also proposed model was tested through the analysis of 20527 papers from the ACL anthology. They found that their proposed model performed well in identifying important and non-important citations with a precision of 90%. They further investigated the proportion of references that were cited in *PLOS ONE* publications in the field of Computer and Information Sciences. 4138 papers and the references cited in them (n=21804) were included in the analysis. Austria was the top country with 13.71% important citations, followed by Sweden (9.77%), Brazil (9.11%), Netherlands (9.06%), and Germany (7.35%). USA (6.83%) was the seventh country and China (4.11%) was the 15th country in terms of the total number of important citations (Hassan et al., 2018).

#### 3.1.1.4 Network analysis

Most studies that have been explained so far had used machine learning approaches in order to identify citation functions or calssify them to different categories. However, Åström (2014) employed co-citation network analysis using VOSviewer in order to visualize the citation contexts of papers citing both Gérard Genette' books and the contexts in which the term "paratext" appeared. Netowork analysis has been used by other researhcers too. For example, Bornmann et al. (2018) used several keyword co-occurrence network analyses to study the impact Eugene Garfield had on subsequent publications.

Åström (2014) analyzed 6853 papers (obtained from WoS) citing Gérard Genette's books and 234 papers citing the "paratext" concept in these books. Most citing papers in both cases came from the area of "literature and language" studies, while less than 20% of citations came from the "humanities" (e.g. philosophy, history, classics, and television). Similar results were found in the co-citation analysis network of journals citing Gérard Genette's books. Co-occurance analyses of keywords, titles, and abstracts of papers citing Gérard Genette's books revealed one theoretical theme (related to literary and cultural studies) and two empirical conceptual themes (related to different contexts, e.g. literary studies, and related to literary works and authors). Co-occurance analyses of keywords, titles, and abstracts in which "paratext" appeared revealed two clusters: one concerned with "hypertext", and one with more diverse terms (e.g. particular genres of text and paratextual elements). Overall, the citation contexts in which "paratext" appeared were mainly focused on empirical concepts, while the Gérard Genette map contained both empirical and theoretical concepts.

Zavrsnik et al. (2016) studied the influence of five pediatric sleeping beatuies on the papers citing them. They used VOSviewer software to create a keyword co-occurrences network based on the titles and abstracts of 926 citing documents (obtained from the WoS). The major finding of this study was that the main contents in the citing papers were associated with the main content discussed in the sleeping beauties. The created network showed that two of the sleeping beauties had a more general and global impact on citing documents than the other three sleeping beauties.

Bornmann et al. (2018) created a keyword co-occurrence network based on the citation context keywords of 59 publications by Eugene Garfield – the recently deceased pioneer of modern citation analysis. The title and abstract information of these papers and their citing papers were obtained from WoS. The citation contexts (n=428) of citing papers were extracted and analyzed. Three co-occurrence networks were created using the VOSviewer software: (1)

a network based on the titles and abstracts of Eugene Garfield's papers, (2) a network based on the titles and abstracts of papers citing Eugene Garfield's papers, and (3) a network based on the citation contexts of papers citing Eugene Garfield's papers.

The co-occurrence network of the citing papers revealed 18 clusters with a variety of bibliometric topics, such as "bibliometric data (e.g. JCR), indicators (e.g. h-index), units of bibliometric analysis (e.g. nations), types of citations (e.g. self-citations), bibliometric methods (e.g. citation networks), and the use of citations in peer review and research evaluation" (Bornmann et al., 2018, p. 432). This shows that Eugene Garfield's papers have been cited by a wide range of publications with various themes. Comparing the three networks demonstrated that semantically there was more similarity between Eugene Garfield's papers and the citation context of papers citing Eugene Garfield than the titles and abstracts of papers citing him. The co-occurrence network of titles and abstracts of papers citing Eugene Garfield revealed two clusters with keywords that didn't occur in the other two networks. "Journal Impact Factor" was the most frequent keyword in the three networks. This study demonstrated the importance of utilizing co-occurrence network analysis of citation contexts in characterizing citations (Bornmann et al., 2018).

#### 3.1.1.5 Other studies

The studies that could not be classified in the sections described above are presented here.

Halevi and Moed (2013) analyzed the contexts of 1150 citations in 32 papers published in the *Journal of Informetrics*. Citations were classified into in-disciplinary and out-disciplinary citations. The distribution of citations in different paper sections was investigated. The XML format of papers were extracted from ScienceDirect and were analyzed at SciTech Strategies (https://www.scitech-strategies.com/). This included the extraction of each citation along with the sentence before and after the citation as well as the sentence in which the citation appeared. This study found that most citations appeared in the introduction section, with 154 out-disciplinary and 125 in-disciplinary citations, followed by the findings section (62 out-disciplinary and 61 in-disciplinary citations), and discussion section (47 out-disciplinary and 55 in-disciplinary citations). However, in the methodology section, more in-disciplinary citations (n=62) were found than out-disciplinary citations (n=24).

Halevi and Moed (2013) explained that the existence of extensive out-disciplinary citations could be due to the multidisciplinary nature of the *Journal of Informetrics*. For example, in the introduction section, multidisciplinary journals such as *Nature*, *American Scientist*, *Scientific American*, and *Science* were cited. The disciplines cited in the introduction section and other sections included "medicine", "social sciences", "physics and astronomy", "science and technology", "business", and "management and accounting". However, these disciplines have been cited with different frequencies across different sections. For example, "mathematics" was cited 5 times in introduction, 1 time in conclusion, and 0 times in other sections. Halevi and Moed (2013) noted that analyzing in-disciplinary and out-disciplinary citations in context increases our knowledge of the relationship between papers published in different disciplines.

Galgani et al. (2015) created a scheme for the automatic classification of legal citations (court cases, decisions, regulations, etc.) in order to characterize the relationship between the cited and citing legal cases. The dataset contained 2027 documents and their citations from the Australian Legal Information Institute. They found that human experts often disagreed on classifying (labeling) legal citations (32% agreement and 68% disagreement). The authors focused on identifying two classes, “distinguished”, and “followed/applied”. Results revealed 460 citations of the “distinguished” class and 3496 citations of “followed/applied” classes. Their classification system identified “distinguished” and “followed/applied” citations with a degree of accuracy of 88.8%. An important finding was that when the citations on which experts didn’t agree were removed, the classification system performance in detecting all classes increased.

In another study, Bakhti, Niu, and Nyamawe (2018) used a dataset of 200 papers and their citation sentences (n=8700) from ACL Anthology. The most frequent functions were “useful” (24.81%), “mathematical” (21.21%), “contrast” (20.68%), “correct” (19.54%), and “neutral” (13.73%). In other words, most studies preferred to use, follow or extend (useful) cited works, and few of them addressed the weaknesses of previous works (correct). The classifier accuracy in recognizing “useful” (62%) was higher than other functions, followed by “mathematical” (61%), “correct” (60%), “contrast” (58%) and “neutral” (58%). This study attempted to improve the performance of citation function classifier by including the author information (author ID, author name, institution, and publication level). Results indicated that the classifier achieved its best performance in identifying citation functions when author information was combined with citations. Their proposed model achieved a degree of accuracy of 62.7%.

Besides the empirical studies mentioned above, Erikson and Erlandson (2014) proposed a “taxonomy of motives to cite” with four categories, including “argumentation”, “social alignment”, “mercantile alignment”, and “data” (see Table 8). They noted that almost all types of paper have the first three categories, while “data” is often seen in papers that analyze previous studies (e.g. reviews and meta-analysis studies).

“Argumentation” has five subcategories and refers to the traditional reasons for citing (supporting a standpoint): “delimitation”, “active support”, “active criticism”, “passive support”, and “further reading”. “Social alignment” has three subcategories and refers to ways in which the citing authors present themselves through the text: “scientific tradition”, “scientific self-image”, and “effort compensation”. Erikson and Erlandson (2014) mentioned that the three subcategories of “social alignment” are different than “perfunctory citations”, because in the former, “by showing affiliations and belongings, the author creates a context for the reader, putting some potential readers off while appealing to others who will read the paper more favorably” (Erikson & Erlandson, 2014, p. 631). The third category, “mercantile alignment”, includes “credit”, “own credentials”, “bartering material”, “self-promotion”, and “pledging”. The fourth category, “data”, refers to the reviews and meta-analysis studies in which cited papers are used as the main source of data analysis. Table 8 shows the taxonomy of motives to cite according to the study by Erikson and Erlandson (2014).

Table 8. Taxonomy of motives to cite (based on Erikson & Erlandson, 2014)

| Categories | Sub-categories | Description |
|---|---|---|
| Argumentation | Delimitation | Refers to citations that are used to clarify what the viewpoints of the citing author are. However, delimitation doesn’t mean that citations are used to criticize or approve the cited work. |
| | Active support | In “active support”, authors cite previous studies as arguments to show that their claims are correct. |
| | Active criticism | Involves providing rationales for a study by criticizing the limitations and weaknesses of previous works (the criticism of general approaches, methodologies, conceptual shortcomings, schools of thought, etc.). |
| | Passive support | Refers to citing because of the reputation of the cited author or cited journal, not for the strength of the arguments presented. |
| | further Reading | Provides audiences with further literature and reinforce the author's position within the scientific community. |
| Social Alignment | Scientific tradition | Disciplines and sub-disciplines are different in terms of scholars' citing behaviors (the number of citations expected, the works considered worth citing, etc.). |
| | Scientific Self-image | Citing to show the tradition to which the authors belong, and the relationship between the citing author and the cited tradition. |
| | Effort compensation | Effort compensation is evidenced when a long or tedious paper that is really not in line with the citing author's manuscript, is cited to get at least something back for his/her efforts in reading the paper. |
| Mercantile Alignment | Credit | Citing to give credit to cited authors. This might also weaken the claims and arguments of the citing author, because, part of the credit for the arguments would be shifted toward the cited authors. |
| | Own credentials | Citing papers to show that the citing author knows the subject very well (e.g. to show that one has the ability to combine various studies). |
| | Bartering material | Citing other authors within the same discipline in order to increase the chance of being cited in return and to expand scientific networks. |
| | Self-promotion | Self-citing with the purpose of drawing others' attention toward one's own publications. |
| | Pledging | Pledging is the kind of citations which is made to impress a journal editor or reviewer (e.g. citing particular theories or authors, or citing the publications in the journal to which the paper is submitted). |
| Data | Review | In "review", a paper is cited to provide readers with an overview of the topic under study. |
| | Meta-analysis | Refers to citing meta-analysis studies in order to create a foundation for new studies. |
| | Text study | In text study, a paper is cited because the paper itself is the subject of study in an empirical study. For example, a cited paper may be analyzed to disclose the opinions and thoughts of its authors. |

### 3.1.2 Manual data processing

Most studies in this section are case studies in which manual data analysis techniques have been employed to investigate how highly-cited classical papers and books have been cited. These studies are more strongly rooted in the traditional citation content analyses than studies using automated data processing techniques. Three of these studies are in the area of organization studies (Anderson, 2006; Anderson & Sun, 2010; Sieweke, 2014); six are in the area of biomedical sciences (Danell, 2012; Liu et al., 2015); and three are in information and computer sciences (Chang, 2013; González-Teruel & Abad-García, 2018).

#### 3.1.2.1 Citation Content Analyses

##### 3.1.2.1.1 Organization studies

Anderson (2006) analyzed the influence of Karl Weick's book on citing studies. Karl Weick's "The Social Psychology of Organizing" is a highly-cited classical book in the area of organization studies. Citations to the book that were received from the top 12 journals for organization studies (e.g. the *Academy of Management Review*, AMR, *Administrative Science Quarterly*, ASQ, and *Organization Studies*, OS) were collected from the Social Science Citation Index (SSCI). Analyzing 578 citation contexts (from 328 citing papers) resulted in 101 distinct concepts, among which Anderson (2006) categorized the frequently cited concepts to 12 categories. These categories represented 67.6% of the citations to Karl Weick's book. The most frequently cited concept was "enactment" (16.6% of citation contexts), followed by "equivocality" (6.6%) and "refutational" (4.7%). One major contribution of this study was its result that regional differences existed in the citing behaviors of scholars. Anderson (2006) demonstrated that the book has been cited for different reasons by the US-based journals versus European-based journals. A reference might be cited with the same frequency across different geographical regions, yet it might be cited for different reasons. For example, 14 (52%) of the 27 refutational citations occurred in OS, a European-based journal.

Anderson and Sun (2010) conducted a study investigating how Walsh and Ungson's (1991) highly-cited paper in the field of organizational memory has been cited. SSCI was searched to identify the papers cited Walsh and Ungson (1991). 496 citation contexts from 301 citing papers were extracted and analyzed manually. The most frequent concepts citing Walsh and Ungson (1991) were "storage bins" (n=170), "general reference to organizational memory" (n=99), "use, misuse, and abuse of organizational memory" (n=90), "definition of organizational memory (whether verbatim or not)" (n=49), and "information processing view of organizations/Example of works" (n=42). The most disciplines citing this work were the "management discipline" (55%), and "information technology" (27%). Only 3.4% of citations were "critical".

The citation content analysis study by Sieweke (2014) investigated the influence of French sociologist Pierre Bourdieu on 352 citing papers published in nine leading journals in management and organization studies. The SSCI was used to collect papers. 63 different concepts were found in the 476 analyzed citation contexts. However, the concepts cited at least 11 times accounted for 66.3% of the total number of cited concepts. 221 (46.6%) of the citations to Bourdieu's work came from three concepts: "capital" (19.3%), "habitus" (13.9%), and "field" (13.4%). Among the different capital forms, "social capital" (47.2%) was the most frequently cited form, followed by "cultural" (18.2%), "symbolic" (17.6%), and "economic capital" (12.8%). Citations were classified according to whether Bourdieu's work was cited in a "limited", "intermediate" or "comprehensive" manner. Results indicated that "at least 50% of the papers in which the three concepts are cited address them in a limited manner" (Sieweke, 2014, p. 535). However, the depth of the citations had increased over time. "Capital" was cited to a limited degree in 62% of the papers, followed by "habitus" (50%), and "field" (50%).

##### 3.1.2.1.2 Biomedical sciences

Siontis et al. (2009) used citation content analysis to find the weaknesses of the two clinical trials mentioned in citing papers. These clinical trials had contradicted the benefits of a treatment technique for coronary artery disease over optimal medical therapy. Siontis et al. (2009) evaluated 87 papers (retrieved from WoS) citing these two trials. Overall, 15 citing papers had reserved positions toward the two clinical trials which were "lack of power, eroded effects from crossover, selective inclusion and exclusion of specific types of patients, suboptimal clinical setting, use of bare-metal stents, suspiciously good results in the conservative treatment arm, and suboptimal outcome choices or definitions" (Siontis et al., 2009, p. 695).

Ramos et al. (2012) analyzed the citation behavior of scholars who had cited two popular ethnobotany papers (Phillips and Gentry, 1993, and c) using content analysis. Citing papers (n=212) were obtained from Scopus. Citation sentences were classified according to their relevance: (1) great relevance (discussed the main ideas presented by the two ethnobotany papers), (2) intermediate relevance (cited quantitative techniques mentioned in the two ethnobotany papers), and (3) minor relevance (not cited the main ideas of the two ethnobotany papers). This study found that the majority of citing papers had "minor relevance" (42.3% of citations for Phillips and Gentry and 56.5% citations for Bennett and Prance), followed by "intermediate relevance" (28.7% of citations for Phillips and Gentry and 38.5% for Bennett and Prance). Less than 20% of citations had cited the main ideas (theoretical contributions) presented by the two papers. Ramos et al. (2012) noted that these results reveal that citing authors barely read the documents they cite, or read them superficially.

Danell (2012) analyzed the citation contexts of 178 papers citing the three highly-cited papers in the area of complementary and alternative medicine (CAM), which were published in the *New England Journal of Medicine* (NEJM), the *Journal of the American Medical Association* (JAMA), and the *Lancet*. For all three papers, 25% of the citing documents were classified as "medicine, general, internal". The top categories citing the NEJM paper were "rehabilitation" (25%), "medicine, general and internal" (24%), "orthopaedics" (22%), and "integrative and complementary medicine" (21%). The main categories citing the *Lancet* paper were "substance abuse" (33%), "medicine, general and internal" (26%), and "psychology clinical" (19%). The main categories in the JAMA paper were "medicine, general and internal" (23%), and "integrative and complementary medicine" (19%). The *Lancet* and JAMA papers were mostly cited by CAM papers, while the NEJM paper was cited by papers outside CAM. The contexts of citations (with a CAM focus) in which the *Lancet* and JAMA appeared were generally of a "positive/confirmatory" nature (40% and 55% respectively). These papers had few "negative/critical" citations (6% and 12% respectively).

The “positive/confirmatory” citation contexts were in many cases relatively short and brief, without going into details. Similarly, “negative/critical” citations were often expressed in a general manner. These citations, however, focused on the general limitations of the cited papers. In contrast, the mixed citation contexts (positive/confirmatory + neutral/empty, positive/confirmatory + negative/critical, neutral/empty + negative/critical) were more detailed and elaborated. “The most important characteristic of the mixed citation contexts is that they not only include positive or confirmatory statements, but also clear objections” (Danell, 2012, p. 317). The NEJM paper has been cited in a clearly more “negative/critical” manner in both CAM focus and non-CAM focus citations (32% and 39%, respectively). Overall, “positive/confirmatory” citations were more frequent in CAM-focused documents, while “negative/critical” citations were more frequent in non-CAM focused documents.

Liu et al. (2015) analyzed 228 citation sentences in 200 papers citing the 2014 Nobel Prize winner John O’Keefe’s work (about the discovery of cell placement). Papers were retrieved from WoS and categorized according to their citation polarity (positive, negative, and neutral). Neutral citations were further interpreted based on citing motivation: “(1) Related work in background or introduction. Introduce the related work with no comments. (2) Theoretical foundation. Concepts, principles, methods, or results which will be used in citing paper. (3) Experimental foundation. Including experimental conditions, processes, environment, and results” (Liu et al., 2015, p. 244). The most frequent terms in citing sentences were “cell placement” (n=76), “hippocampus” (n=74), and “environment” (n=55). Most citations were “neutral” (n=204). Among the neutral citations, “related work” had the highest frequency (n=114), followed by “theoretical foundation” (n=49), and “experimental foundation” (n=41). Only 24 positive citations, but zero negative citations, were found among the 228 citing sentences.

Cristea and Naudet (2018) conducted a citation content analysis to identify how Leucht et al. (2012), an influential paper in psychology, was cited in the literature. This influential paper concludes that the effect of psychiatric drugs was somewhat like the effect of drugs in general medicine. 120 citing papers retrieved from Scopus were analyzed. 53% of the papers had cited “Leucht et al.’s paper to justify a small or modest effect observed for a given therapy” (Cristea & Naudet, 2018, p. 230). 60% of the papers had cited the paper to claim that no substantial difference in treatment effectiveness exists between psychiatry and general medicine. Regarding the cited conditions, 35% of the papers had cited the paper in a general context, 28% for affective disorders,16% for psychosis and other reasons such as referring to addiction and attention deficit hyperactivity disorder (ADHD). 87% of the papers had cited Leucht et al. (2012) to make a comment about the effects of drugs and psychological interventions. Among treatment categories, 34% of papers had cited Leucht et al. (2012) without mentioning any specific treatment category, followed by various treatments (31%), antidepressants (23%), and antipsychotics (15%).

##### 3.1.2.1.3 Information and computer science

The study by Chang (2013) examined the 1963 and 1965 editions of de Solla Price’s book “Little Science, Big Science” (LSBS) – a landmark publication in scientometrics – to identify the cited concepts and citation functions and to investigate differences between natural sciences (NS) and social sciences and humanities (SSH) disciplines. 908 papers citing de Solla Price’s book were obtained from WoS, and their citation contexts (n=1142) were analyzed. Ulrichs Global Series Directory, the Library of Congress (LCC) and the Dewey Decimal Classification (DDC) were used to classify cited papers as NS or SSH papers. Cited concepts were extracted from the text surrounding in-text citations. “Science growth patterns” (16.1%), “scientific communication” (15.7%), and “scientific productivity” (12.9%) were the top concepts in NS. “Scientific communication” (16.2%), “science growth patterns” (13.2%), and “scientific productivity” (10.8%) were the top concepts in SSH.

Citation contexts were classified according to one of the following citation functions (Chang, 2013, p. 540):

“1. Background information: LSBS offered information to help readers understand the background of research questions.

2. Comparison: LSBS was used as a basis of comparison in the citing paper.

3. Definitions: The definition of certain concepts originated from LSBS.

4. Evidence: LSBS served as evidence to support the citing author’s statements.

5. Figures: Statistics or other quantitative data.

6. Further reading: LSBS was suggested as additional reading.

7. Methods: The methods and data processing used in LSBS.

8. Related studies: LSBS was among the previous studies related to a specific topic.

9. Supplement/explanation: The content of LSBS gave additional information on, or was used to explain the reasons for something.

10. Terms: The citing paper used terms contained in LSBS.

11. Views: Price’s views contained in LSBS”.

The most frequent functions in NS were “evidence” (20%), “related studies” (16.9%), “background information” (11%), “terms” (11%), and “views” (10.9%). The most frequent functions in SSH were “evidence” (22.3%), “related studies” (21.5%), “views” (14.5%), “background information” (11.7%), and “terms” (7.6%) (Chang, 2013).

In a recent study, González-Teruel and Abad-García (2018) conducted a citation content study to investigate the influence of Elfreda Chatman’s theories (information poverty theory, IPT, life in the round theory, LRT, and normative behavior theory, NBT) on citing papers. The full text of 332 citing papers were obtained from the WoS and Scopus. Most of the citing papers were in “social sciences” (39% for IPT, 29.4% for LRT, and 66.7% for NBT), “computer science” (31.7% for IPT, 47.1% for LRT, and 33.33% for NBT), and “medicine” (19.5% for IPT, 11.8% for LRT, and 33.3% for NBT). The concepts citing Chatman’s theories were as follows:

1) Library and Information Science related terms, such as information, information seeking behavior, and information sources.
2) General terms used to express the three theories, and previous studies that were used as a basis for the three theories such as virtual communities and feminist booksellers.
3) Terms that were not core concepts, but frequent, such as outsider and insider, small world, social norms, worldwide, and social types.

McCain and Salvucci (2016) conducted a content analysis of 574 citation contexts from 497 papers citing Frederick P. Brooks’s book, “The Mythical Man-Month”. The citations to Brooks’ book were obtained from ISI files (Dialog files). Most papers were cited from “software engineering” (n=139), “computer science” (n=137), “management and industrial engineering” (n=61), “other sciences” (n=31), “electrical engineering” (n=23), and “information systems” (n=21). The concepts for which the book was cited were “generalia” (general description of the book), “project management issues”, “building the system” and “other concepts”.

#### 3.1.2.2 Other studies

Camacho-Miñano and Núñez-Nickel (2009) reviewed 139 papers to answer the question as to why authors prefer to select some references over others. They proposed a model with three phases in selecting references: (1) “external limitations”, (2) “functional choice”, and (3) “preferential selection”. Phase 1 refers to several external limitations in accessing scientific papers (some papers are difficult to access), and restrictions that exist in reading and understanding papers that are written in foreign languages. In the “functional choice” phase, citing authors would objectively classify and select papers according to the functions they have for their paper. These functions are explained in Table 9. When authors obtain and select all useful papers according to their functions, the third phase (preferential selection) takes place. In the third phase, scientific and non-scientific criteria would be employed to select the final list of references. Camacho-Miñano and Núñez-Nickel (2009) classified the non-scientific reasons for citing into several subjective prejudices, including author, journal, and paper (e.g. number of pages) among others. For example, a paper published in a prestigious journal or by a reputable author may be conceived as more relevant and reliable to the citing paper, and accordingly would be selected and cited (Camacho-Miñano & Núñez-Nickel, 2009).

Table 9. Functions of citations in the text.

| Function | Explanation |
|---|---|
| Conceptual | Citation is useful for showing concepts, definitions, or interpretations, or for substantiating a statement or an assumption. |
| Operational | Citation contributes additional information, data, a point of comparison, a theoretical equation, or methodology. Results of the citing paper furnish a new interpretation/explanation of the data of the cited source, a methodology, or formulation of research problems. |
| Organic | Parts of relevant literature that are influential, essential, basic; descriptions of other relevant work. The results of the citing paper prove, verify, or substantiate data or interpretation of the cited source. |
| Perfunctory | Citation that is casual, unusual, neutral, or made with reservations, or for ceremonial purposes. It is included as a note or with no clear indication of reasons. |
| Evolutionary | Historical background; citations are mentioned in the introduction or discussion as part of the history and the state of the art. |
| Juxtapositional | Additional information that is supplementary or illustrative. |
| Confirmative | Cited source is positively evaluated, is of critical importance to results. |
| Negational | Cited source is negatively evaluated: partial or total negation; for example, when a theory or method is not applicable or not the best one, the citation is made with criticism and another treatment is proposed by author. It could be with the aim of correction, discussion, or disclaimer. |
| Others | Alerting readers to forthcoming work, anticipated value, or new research. |

Source: Camacho-Miñano and Núñez-Nickel (2009, p. 757)

In a recent study, Lin (2018) conducted a study how "essential" versus "perfunctory", and "confirmative" versus "negational" citations were distributed across six sub-disciplines in the humanities (Chinese literature, history, and art) and the social sciences (sociology, economics, and psychology). Papers were retrieved from the Taiwan Humanities Citation Index and Taiwan Social Sciences Citation Index. Six people manually annotated 25617 in-text citations from 360 papers (60 for each discipline). The authors designed a two-dimensional coding scheme based on Moravcsik and Murugesan (1975) citation function classification. The dimensions were "essential/perfunctory" and "confirmative/negational". The citation functions were as follows: "essential-concept-confirmative", "essential-concept-negational", "essential-factual-confirmative", "essential-factual-negational", "essential-methodology-confirmative", "essential-methodology-negational", "perfunctory-confirmative", "perfunctory-negational" (Lin, 2018, pp. 799-800).

Their findings showed that 97.13% of citations were "confirmative" and only 2.83% were "negational", which accords with the unbalanced results in previous studies. The humanities had more "negational" citations than social sciences (4.17% vs. 1.7%). History had 9.1% "negational citations" (see also Lin, Chen, & Chang, 2013). More "perfunctory" citations were found in the humanities than in the social sciences. History with 52.2% in the humanities, and economics with 43.9% in the social sciences, had the highest ratio of "perfunctory" citations. Overall, "perfunctory" citations in the six subject categories analyzed accounted for less than 50% of citations. Almost all papers (except two) had at least one "essential-confirmative" citation. Chinese literature had the lowest number of "perfunctory-confirmative" citations. Chinese literature with 1.8% and economics with 1% had the lowest number of "negational" citations. It was found that "the humanities and social sciences as two groups were not significantly different in authors' general negational behavior and in essential negation, but were significantly different in perfunctory negation" (Lin, 2018, p. 811).

## 3.2 Citer motivation surveys or interviews

In addition to citation content/context analyses, surveys and interviews have also been used as two other important approaches in exploring citing behaviors of scholars (Bornmann & Daniel, 2008). In surveys and interviews, the citing authors are being asked to express their reasons for citing the literature. Figure 2 explains the components of citer motivation surveys and interviews.

Harwood (2009) mentioned that textual analysis has several weaknesses for content/context analysis purposes, because (1) sometimes the motivation of the citing author is not explicitly apparent even after reading the text, and (2) a successful text analysis demands the researchers to have an in-depth knowledge of the document's research area. Willett (2013) revealed that there was a limited level of agreement between citing authors' judgments of their own reasons for citing and the judgements of independent readers. Willett (2013, p. 151) showed that "different readers understand the contexts of the same citations in very different ways". The author noted that readers (experts) may be able to easily identify possible reasons for citing, although they are unlikely to correctly perceive authors' reasons for citing. Semi-structured interviews with the citing authors may overcome some of the limitations of content/context analyses by enabling citing authors to explain the functions of the citations they have used in their own papers (Harwood, 2009).

Harwood (2009) conducted semi-structured interviews with six computer scientists and six sociologists to identify the functions of citations in their papers or book chapters published recently. This study found eleven citation functions (see Table 10). The most frequent function was "position" in sociology (46.2%) and "signposting" in computer science (26%). "Position" (24.8 %) was the second most frequent function in computer science. "Engaging" was the most frequent function in sociology (17.2%), while it occurred with low frequency in computer science (0.88%). Harwood (2009) posited that it is mainly the disputational nature of sociology that leads to a higher proportion of engaging citations than in computer science. The third, fourth, and fifth top functions in both fields were "supporting" (20.2% in computer science and 15.28% in sociology), "credit" (15.11% in computer science and 9.24% in sociology), and "building" (6.66% in computer science and 6.39% in sociology), respectively. Harwood (2009) also showed that a citation may have more than one function: "over half of the citations in both fields were said to have more than one function" (Harwood, 2009, p. 495).

Table 10. The eleven citation functions identified by Harwood (2009)

| Citation functions | Definition and application |
|---|---|
| Signposting | Signposting is similar to "further reading" in Erikson and Erlandson (2014). This function directs readers to other sources to (1) inform readers of other relevant and interesting papers, (2) to encourage readers to read more about the argument, and (3) to save space. |
| Supporting | This function refers to the citations that are used to justify their topic, methodology and claims. |
| Credit | This function points to citing others in order to pay respect and acknowledge their own ideas, findings, methods, etc. (self-defense). |
| Position | Position refers to citations that are used to draw attention toward different viewpoints, clarify cited authors' perspectives in detail, and show the development of cited authors' perspectives over time. |
| Engaging | Engaging citations are seen when a cited work is criticized in a mild or harsh way. |
| Building | This function appears when the methods or ideas of the cited document are used and developed further. |
| Tying | This function refers to the citations that associates the citing authors with other researchers' methodological approaches, schools of thought, paradigms, and the disputes on specific scientific topics. |
| Advertising | This function is used to make readers aware of the earlier works of the citing author or other researchers. |
| Future | Future citations point out to the citing authors' future research plans. |
| Competence | Citations of this type are used to show the citing author's knowledge of the main literature in the discipline and their ability to conduct research. |
| Topical | Topical citations are employed to show that the citing authors and their studies are concerned with the newest and most important topics in the field. |

As part of his interview-based study on sociologists and computer scientists, Harwood (2008b) investigated the impact of the publication outlets, in which authors' works appeared, on authors' citation behavior. The analysis of interviews revealed several themes, including "less specialized outlets", "the effect of co-contributors' citations", "parsimonious citing policy", "festschrifts", "outlet's favored research paradigm", "audience location", and "publication speed". Harwood (2008b) showed that less specialized outlets, such as books for people with general knowledge of the topic, made authors cite references that provided general information for people outside the area. Even when the publication was a journal paper, some basic references were used to provide some background knowledge and make the text understandable for a broad range of audiences.

Results revealed that occasionally, publishers and gatekeepers had their own citation preferences and instructions. For example, some publishers asked authors to mention the first name of cited authors along with their surnames in textbook/dictionary entries, in order to make them "less intimidating and 'more personal' for the undergraduate reader" (Harwood, 2008b, p. 257). In journals, however, the policy was often to only mention the surname of cited authors. Sometimes co-authors removed or added some references that were already cited by other co-authors. For example, in one case an influential researcher in the field was not cited to such an extent that more citations to her were later added as an "acknowledgement of the influence of her work" (Harwood, 2008b, p. 258). Parsimonious citing policy was another theme that emerged in the interviews, whereby co-editors preferred to recommend the citation of their own papers (Harwood, 2008b).

Another theme that emerged in this study was the "commemorative publication outlet", which affected the way participants cited. For example, although the policy made them cite by mentioning the surnames of the cited author, in the case of citing an honoree, his/her first name was also included, "because this is the way people knew him" (Harwood, 2008b, p. 259). The "commemorative publication outlet" also explained a case in which the publisher asked the contributor to "go easy" on a paper that wasn't related to the theme of the paper, but which was cited by other co-contributors. Another theme in the study was the “outlet’s favored research paradigm” which referred to circumstances where citing the literature has been based on the favorite paradigm (quantitative or qualitative) of the target journal. “Audience location” was another theme which was evidenced in some cases when the citing author had cited a source mainly because the predominant audiences of their paper were from a specific geographical area. The “outlet’s space restrictions” was another factor that contributed to favoring some references over others. This was the main reason for citing some sources only, in order to “direct readers to relevant sources for fuller explanations” (Harwood, 2008b, p. 261), a phenomenon known as “signposting citation”. The other theme, “publication speed”, mainly led to incomplete references. One interviewee explained that some of his cited references were “not as up to date as they could be … [because of] … the time lag between his writing the chapter and the publication of the volume” (Harwood, 2008b, p. 261).

Harwood (2008a) also investigated the reasons for which the names of cited authors were mentioned in different formats. The themes that emerged from this study were “stylistic elegance”, “stylistic variation and informality”, “other stylistic preferences: integral and non-integral citations”, “making the text accessible”, “revealing the citer’s politics”, “acknowledging seminal sources”, “responding to reviewers’ requests”, and “unconscious, arbitrary, and/or inexplicable motivations: accident, not design”. “Stylistic elegance” referred to cases when the name of the cited author was mentioned to make the sentences read better. “Stylistic variation and informality” referred to self-citations where authors preferred not to mention their own names, and rather referred to themselves by saying “they” or “the authors”. “Integral citations” were those citations in the Harvard system where the first names of cited authors were used (Harwood, 2008a).

“Non-integral citations” were those in which the name of the cited author appeared in parentheses. “Making the text accessible” referred to citations where the first names of the cited authors were mentioned as an influence of the previous experience in writing undergraduate textbooks. In this regard, one of the participants claimed that “textbook publishers like authors to include the first name of cited authors to make the writing less intimidating for the undergraduate readership” (Harwood, 2008a, p. 1008). Other participants, however, reported some cases when first names were accidentally used in their writings. “Revealing the citer’s politics” was another theme, which referred to the cases when the first names were mentioned in order to make the gender of the cited author recognizable or to demonstrate the close relationship between the citing and cited author. One of the participants said that mentioning the first name was “a gesture of affectionate friendship” (Harwood, 2008a, p. 1009). Additionally, according to one of the participants in this study, if the cited reference was from another discipline and unknown to many audiences, using the first name would help them know and remember the work. Using the first name was also evidenced in “acknowledging seminal sources”, when the writing included a historical overview of the field by citing the fathers of the research area. In some cases, in order to make it more obvious that the referees’ requests were carefully taken into consideration, the first names of the cited authors were intentionally or unconsciously mentioned in the citing document by the citing authors. However, it should be noted that authors’ motivations for mentioning the first name of the cited author were in many cases “unconscious” and “arbitrary” (Harwood, 2008a).

Tang and Safer (2008) asked 50 psychologists and 49 biologists to rate the importance of the cited references in their own publications, and to explain their reasons for citing them. This study further explored which textual features of the cited references (frequency of citation, citation length, location of citation in the text, and citation treatment) could be used to predict citation importance. Results indicated that citations were rated fairly important. 64.7% of references in psychology and 50.9% in biology were cited in the introduction section. The most frequent reasons for citing were "general background" (37.3%), "conceptual ideas" (31%), and "methods and data" (13.4%). The citing reasons in terms of different paper locations are presented in Table 11 (Tang & Safer, 2008).

Table 11. Distribution of authors' citation reasons for references occurring in various locations

| Citation reasons | Introduction (%) | Method (%) | Result (%) | Discussion (%) |
|---|---|---|---|---|
| General background | 46.0 | 16.6 | 26.9 | 30.9 |
| Conceptual idea | 36.1 | 17.8 | 28.3 | 36.1 |
| Method and data | 6.4 | 47.6 | 26.6 | 7.0 |
| Correctness of method/result | 6.8 | 14.6 | 15.7 | 14.5 |
| Dispute or correct it | 2.5 | 1.0 | 1.9 | 2.4 |
| Suggest limitations | 0.5 | 0.9 | 0.0 | 2.4 |
| Future studies | 1.6 | 1.5 | 0.5 | 6.7 |

Source: Tang and Safer (2008, p. 260)

Authors' disciplines, gender, ranks, and affiliations did not influence citation importance ratings. Citations with the following features were rated as being more important than others:

- Recurring citations with a higher frequency
- Citations with a longer length
- Citations having received more in-depth treatment (e.g. citations in method)
- Citations with conceptual ideas, methods, and data functions
- Citations whose authors were known to the citing author
- Self-citations when they were cited for "conceptual ideas", "method", and "data" reasons

In terms of citation treatment or depth, only 4.1% of cited references in both fields were quoted or discussed multiple times, 3% of references were thoroughly discussed, more than 80% were not precisely mentioned, and 10% were barely mentioned in the text. References whose authors were known to the citing author received greater in-depth treatment and higher citation frequency. References that were cited for the reasons of "conceptual idea" and "method and data" were likely to receive more in-depth treatment than "general background". The citation treatment analysis further indicated that cited references, whose first authors were senior researchers, received more in-depth treatment than those in which the first author was a junior researcher (Tang & Safer, 2008).

Milard (2014) conducted semi-structured interviews with 32 French chemists in order to characterize the type of social relations between themselves and the authors they had cited in their papers. 32 chemistry papers from the French chemists published in prestigious international journals indexed in WoS were analyzed, along with their cited references (n=1410). The authors were asked to address the history of their paper, such as its origin and relation to the cited papers. Several social circles were identified in this study: “co-authors”, “close friends”, “colleagues”, “invisible colleges” (have had discussions by e-mail), “peers”, “contactables” (unknown people but co-authors of known authors), and “strangers” (totally unknown). Results revealed that purely social references were rare (20 out of 1410 references).

This study found that 25.2% of cited authors were “unknown” and 74.8% were “known” to citing authors. However, knowing the cited author was not always the reason for citing. “Personal relationship” was a major criterion when authors had to choose among several cited references. Some types of distances were found in relation to why the cited authors were unknown to citing authors: (1) “social distance” (in earlier publications, the cited author was less known), (2) “geographic distance” (citing and cited authors living in far distant geographical areas), (3) “disciplinary distance” (not in the same specialty), and (4) “professional status distance” (lesser known authors, because they were not senior scholars) (Milard, 2014).

Clarke and Oppenheim (2006) conducted a survey investigating the citing behavior of 65 postgraduate students in Loughborough University Department of Information Science. Results showed that the most frequent reasons for citing was to “support their own argument” (95.4%), and because the cited papers were “up-to-date” (95.4%). “Giving positive credit to related works” (94.4%) and “persuading readers” (92.3%) were other most frequently mentioned reasons to cite. 80% of students said they had cited to “criticize” other works. 56.9% of students disagreed that they had cited because the cited author had been their tutor. “Well-known researchers” seemed to be another reason for citing for 70.8% of the participants. Clarke and Oppenheim (2006) noted that their results showed that students believed they more often based their decisions to cite on the “relevance” and “importance” of the cited works than other (non-scientific) reasons.

Another survey study by Thornley et al. (2015) investigated the extent and the reasons for which cited references were regarded as authoritative and trustworthy. Thornley et al. (2015) distinguished their study from previous ones by pointing out that their study investigates scholars’ citation behavior in the context of trust. As such, participants were asked why they had trusted a particular reference rather than why they had cited it. This study found that the reasons for citing were multi-dimensional and depend on a variety of factors. According to this study, citation counts were considered as one indicator of trust and authority. However, the authors noted that citations could not be regarded as the sole measurement of quality. The most popular reason for citing was knowing the cited authors (24.16%). Cited authors were considered as more trustworthy if citing authors had already met them or were familiar with their works. In 15.1% of cases, papers were cited because they were classical and seminal works. This study also found that papers that did not refer correctly to seminal sources were not trusted. “Trusted sources” were mentioned as an indicator of quality and were used as a criterion for citing. It was found that citing was sometimes the result of trusting co-authors and discussion on choosing references, or they were sometimes suggested by editors or reviewers.

This study also explained whether the reasons for trusting references supported or refuted the normative theory (citing due to the quality of content) or social-constructivist theory of citation (citing for social and political reasons). Thornley et al. (2015) indicated that acknowledging the quality and influence of content was a central reason for citing, but other reasons were also taken into consideration (e.g. journal reputation). Normative reasons accounted for almost 19% of the reasons to cite. However, in many cases the normative and social-constructivist reasons for citing seemed to be linked.

Thornley et al. (2015) further demonstrated that citations had different functions and levels of importance for the citing papers, some of which might not be explained by the normative and social-constructivist theories. They noted that the normative theory explained citation functions better than the social-constructivist theory, while the latter provided a better understanding of how trusted social networks influenced citing. Thornley et al. (2015) demonstrated that in nearly all cases, researchers had carefully read the documents they cited, which seems to be in line with the theory of Nicolaisen (2007), who proposed that authors avoid careless citing. However, this result is contrary to the findings of other studies (e.g. Ramos et al., 2012; Wright & Armstrong, 2008).

## 4 Summarizing the empirical results

In this study, we reviewed the studies from 2006 to 2018 on the relationship between citing and cited documents in order to describe the reasons for citing. The studies were classified into (1) citation content/context analyses and (2) citer motivation surveys or interviews. We also explained some of the technical developments, such as the machine-readable format of full texts, which had an enormous influence on the methods of conducting citation content/context studies in the last decade. These technical developments facilitate comprehensive studies that had not previously been possible.

Our literature overview demonstrates that the reviewed empirical studies had used different methods, techniques, algorithms, schemes, and datasets to classify citation functions, and to make classifying citation functions automated. As such, different schemes with small or large numbers of categories have been proposed to classify citations. Some citation functions used in previous studies had (to some extent) the same meanings and definitions; however, different researchers had used different names for them. The citation function classification schemes indicated that citing motivation is a multi-dimensional phenomenon, and scholars cite the literature for a variety of scientific and non-scientific reasons. Possibly, rather than classifying citations into many functional categories (every study introduces a new classification system), using some general but exclusive functions would lead to comparable and (perhaps) more reliable results.

For example, a classification scheme could contain “background”, “methodological”, “constant/comparison”, “negational”, and “perfunctory” citations. To assign citations to more general categories, they could be classified as “influential” (or useful) versus “non-influential” (or perfunctory), as evidenced by some previous studies (based on Hassan et al., 2018; Pride & Knoth, 2017; Valenzuela et al., 2015; Zhu et al., 2015). Citations could also be classified in terms of their relevance into “great relevance” (cited the main ideas presented in cited papers), (2) “intermediate relevance” (cited the techniques mentioned in the cited papers), and (3) “minor relevance” (not cited the main ideas or the techniques in the cited papers) (see Ramos et al., 2012). However, narrowing citation functions (and reasons) to more general categories may not represent all of the functions that citations could potentially have. For example, classifying citations in terms of their influence (importance) or relevance would (possibly) ignore the functions described by Harwood (2009) or Camacho-Miñano and Núñez-Nickel (2009), such as topical, future, and conceptual functions.

Most empirical studies used machine learning and computational techniques in order to test their proposed citation function classification schemes on different datasets. Such studies often employed machine learning experiments in order to replicate human annotations (Teufel et al., 2006). Studies have mainly attempted to manually create citation function classification schemes that would show high accuracy when they are used to automatically classify citations (Bertin, Atanassova, Sugimoto, et al., 2016). These "studies have focused on three issues: manual annotation of a corpus using different category and function schemes with different approaches; automatic labeling of the corpus from training data, and addressing the problem of defining features that will achieve the best results in citation classification" (Hernández -Alvarez & Gomez, 2015, p. 335). The results of these studies demonstrated that the proposed classifiers for detecting citation functions achieved different levels of accuracy. In some cases, adding or removing one feature resulted in a lower or higher degree of accuracy in detecting citation functions. The results are, however, not comparable between the studies due to the diversity of the classification schemes, methodologies, and datasets. There is no consensus regarding a single standard scheme which could be used for most disciplines or even for one certain discipline (Hernández -Alvarez & Gomez, 2015). Here, it seems necessary that the studies are better rooted in each other.

Detecting citation functions, using either automated or manual data processing, is often a difficult task for a variety of reasons, mainly because the citations' purposes are not always explicitly mentioned by the citing authors (Hernández -Alvarez & Gomez, 2015). In some cases, the results are inconsistent with each other, which makes it difficult to decide which feature should be included in order to produce more accurate classifiers. For example, the study by Pride and Knoth (2017) showed that abstract similarity could be considered as a feature for detecting citation functions, while Valenzuela et al. (2015) showed contrary results. Some features, however, have been found to be more important in detecting citation functions than others, such as the "frequency with which a citation was mentioned in the citing article" (Pride & Knoth, 2017; Valenzuela et al., 2015; Zhu et al., 2015). Overall, in order to better understand the relationship between citing and cited documents, it is essential that the "citation context", the "semantics and linguistic cues" in citations, "citation locations" be analyzed within the citing document, alongside "citation polarity" (negative, neutral, and positive).

A variety of citation functions and reasons for citing were found in the reviewed empirical studies. Results showed that some disciplines may have more functions and functions that differ from other disciplines, possibly because of the differences in scholars' citing behaviors in different disciplines. For example, the study by Sula and Miller (2014) revealed that linguistics showed the most "positive" citations and philosophy had the most "negative" citations. Another study found more "negational" citations in the humanities than social sciences (4.17% versus 1.7%) (Lin, 2018).

Our review showed that some citation functions were in line with the normative and social-constructivist theories of citing. Sometimes, there is a mixture of both. For example, the surveys and interview-based studies demonstrated that personal/professional relationships with the cited authors was one prominent reason for trusting and citing them.

Thornley et al. (2015) showed that knowing cited authors was a popular reason for trusting and consequently citing their scientific works (24.16%). Milard (2014) found that 74.8% of cited authors were known to citing authors. In contrast, "negational" citation was found however to be rare among most studies (e.g. Anderson, 2006; Teufel et al., 2006), possibly because they might "jeopardize a friendship" (MacRoberts & MacRoberts, 1984, p. 92) or might be considered as "potentially politically dangerous" (Teufel et al., 2006, p. 105).

"Use" was a frequently used and popular function in the classification scheme of several studies. For example, Jha et al. (2017) indicated that the most frequent citation function was "use" (17.7%), and that the accuracy of their proposed classifier in recognizing "use" (60%) was higher than that of other functions. "Use" was found to be mentioned quite frequently in most sections, specifically methods and introduction sections (Bertin, Atanassova, Sugimoto, et al., 2016).

Several studies found that only a small percentage of citations were actually "influential" (Pride & Knoth, 2017). For example, Valenzuela et al. (2015) and Pride and Knoth (2017) showed that approximately 14% of the citations were "influential"; Zhu et al. (2015) reported a percentage of only 10.3%. Hassan et al. (2018) found that the number of important citations was very low for most countries. For example, Austria, with just 13.71%, had the highest number of important citations, followed by Sweden (9.77%). Some functions are relevant in most disciplines, while others are rather rare (such as influential and negative citations). Pride and Knoth (2017) argued that much larger datasets should be analyzed in order to create and test classifiers that can accurately recognize rare citation functions.

The mentioned studies have used a range of features to classify citations to influential and non-influential citations. Valenzuela et al. (2015) utilized features such as the total number of times a paper was cited, author overlap, abstract similarity, and citation locations (e.g. a citation in the methods section was considered an important citation, or a citation in the related work section was considered incidental). Zhu et al. (2015) proposed a model to detect influential or non-influential papers, using a wide range of features: (1) count-based features (e.g. the occurrences of each reference), (2) similarity-based features (e.g. similarity between the title of a cited reference and the title of the citing document), (3) context-based features (e.g. a list of words to classify the citation context), (4) position-based features (e.g. the location of a citation in the paper), and (5) miscellaneous-based features (e.g. the number of citations a paper received). Pride and Knoth (2017) examined the accuracy of three features (the total number of times a paper was cited in the citing document, author self-citation, and abstract similarity) in predicting influential citations. Hassan et al. (2018) utilized "context-based" features, "cue word-based" features, and "text-based" features to recognize important citations. Context-based features represented the total number of citations in different paper sections. A paper was labeled as important, if it was cited multiple times in the citing paper. Citations in the methods section were considered more important than those in other sections. The cue word-based features represented the words that had been used in different contexts: related work citations, comparative citations, using the existing work, and extending the existing work. The text-based features represented the similarity between the abstract of the cited paper and the text of the citing paper (Hassan et al., 2018).

Teufel et al. (2006) were among the pioneer scholars who distinguished between citation polarity and citation purpose. Citation polarity could be used to more effectively assess the actual impact of the cited works on citing works (Hernández -Alvarez & Gomez, 2015). Citation polarity determines the attitudes of the citing authors toward the cited paper, which could be favorable (positive), unfavorable (negative) or neither of these (neutral). However, citation purposes refer to citers' reasons and motivation in citing documents (Hernández-Alvarez et al., 2017; Jha et al., 2017). Positive citation contexts may be indicative of whether the cited author's arguments, claims, thoughts, methodologies, etc. are supported or confirmed. Negative citation contexts may dispute claims, show points of disagreement and weaknesses of previous works, etc. (Sula & Miller, 2014). Investigating the positive, negative, and neutral attitudes of citing authors toward a cited paper discloses the overall attitude of the scientific community about the cited paper (Jha et al., 2017). Some studies have mapped citation polarity to citation function in order to achieve more accurate classification schemes and better results (e.g. Hernández-Alvarez et al., 2017; Jha et al., 2017).

Most citation content/context studies have tried to overcome the methodological weaknesses of previous studies, by proposing new classification schemes, new methods and features for detecting citation functions, or new computational techniques. Bertin, Atanassova, Sugimoto, et al. (2016) noted that most of the previous studies lack a clear operationalization that would allow for large-scale automated analyses. They proposed a natural language processing approach for identifying the functions of citations, considering the location of citations. They assessed whether these linguistic patterns varied according to citation locations in the text. They noted that taking into account both the rhetorical structure of citations and their location would lead to a better understanding of how citations are used in scientific papers (Bertin, Atanassova, Sugimoto, et al., 2016). Jha et al. (2017) mentioned that an accurate classification scheme should contain both the citation purpose and citation polarity for the same sentence, which could be helpful for demonstrating the scientific impact of papers (Jha et al., 2017). Hernández-Alvarez et al. (2017) noted that the best alternative to citation counts as an indicator of scientific influence or impact should be the consideration of "citation function", "citation polarity", "citation frequency" and "citation location", altogether.

A wide variety of techniques have been proposed for analyzing the citation context of papers, most of which take advantage of machine learning experiments. Technical advances have also made it possible to conduct large-scale studies on large datasets in order to gain a better insight into citation motives with results that are less likely to be biased (Bertin, Atanassova, Gingras, et al., 2016). In addition, in recent years, several new databases have emerged which have facilitated citation context studies by providing the machine-readable formats of papers. For example, Microsoft Academic has recently been used by some scholars in order to conduct scientometric studies. In contrast to traditional search engines such as Google Scholar, Microsoft Academic allows automated searching through its Applications Programming Interface (API), and provides better performance when it comes to showing the citation context of papers and other information (Kousha, Thelwall, & Abdoli, 2018; Thelwall, 2018a, 2018b).

We also reviewed several recent studies that had used semi-structured interviews to explore citation motives. One advantage of interviews is that a non-specialist interviewer can discuss and ask questions in order to comprehend the citing behaviors of citing authors (Harwood, 2009). However, interviews have their own limitations. For example, to a certain extent, the recall issue and memory limitations of participants make it impossible to truly determine why they have cited a particular paper (Anderson, 2006; Harwood, 2009). One suggestion for remedying the recall issue is to investigate the papers that were published only a few months prior to the time when a citer motivation survey or interview is being conducted, since citing authors are more likely to remember their reasons for citing (Willett, 2013).

Thornley et al. (2015) showed in their pilot interview with three researchers that participants accurately remembered their citation motivations for recently published papers. Another limitation of interviews is fatigue, as a result of which some respondents may not be willing to elaborate on their responses. Another issue is the possibility that the responses of interviewees might in some cases be self-serving (Harwood, 2009). However, along with other methods such as citation content/context analyses, semi-structured interviews make a better understanding of the citation process possible (Anderson, 2006).

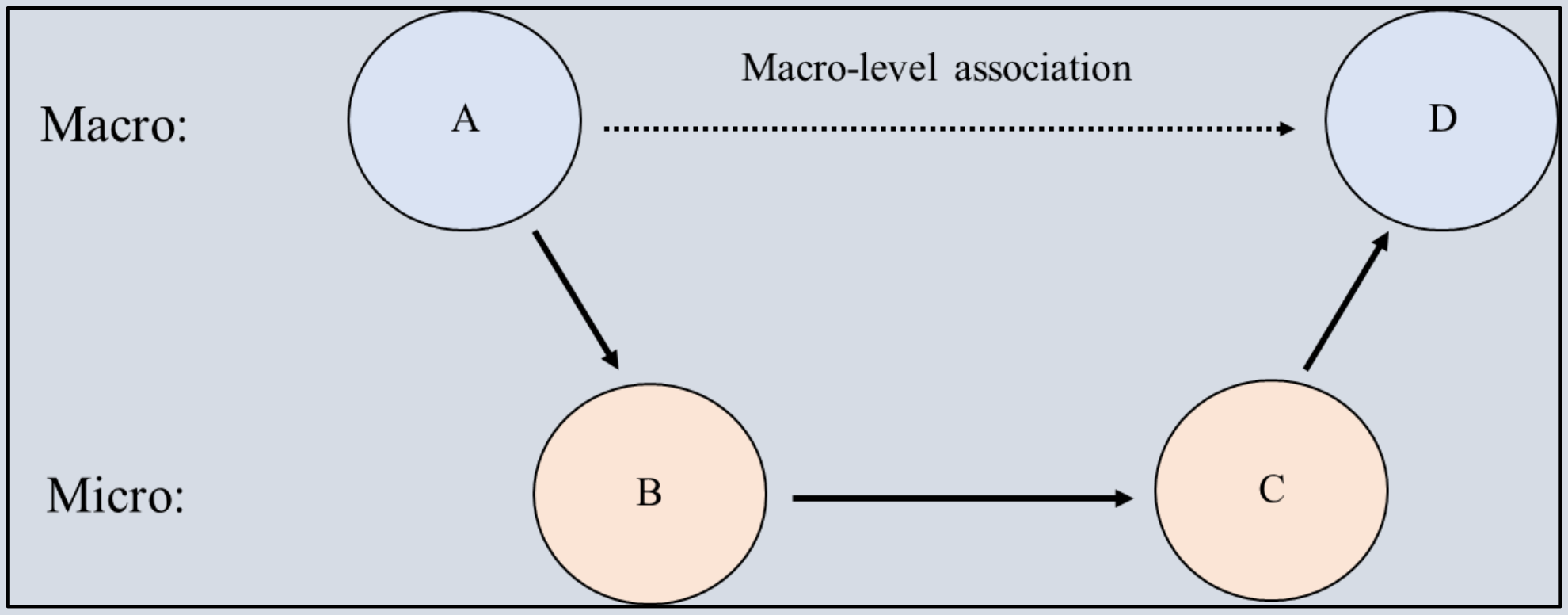

A: Norm in science to cite prior research (on which research stands)
B: Attitudes of single authors (
authors’ **belief** they should cite papers from which they quote text;
authors’ **belief** they should cite the papers of reputable researchers;
authors’ **desire** to have more citations for own papers – and increase self-citations;
authors’ **opportunity**: set of papers which can be possibly cited
…)
C: Actions of single authors (
cite papers **one** or **multiple times** in a document;
discuss cited papers **extensively**;
cite papers in **a long list** with other papers; …)
D: Social phenomenon (
citation impact differences between entities depending on which actions are counted: simple citations; multiple citations in single papers; in-depth discussions of papers …)

Figure 4. Macro-micro-macro model for explaining the relationship between the scientific norm to cite and evaluative results for entities (researchers, institutions etc.) – depending on the actions of citing authors.

Source: adopted and adapted from Hedström and Ylikoski (2010, p. 59)

Based on the results of the studies presented in this review, we propose a model for explaining the way from the norm in science of citing previous literature, actions of citing authors, and the results of evaluative bibliometrics (depending on these actions, see Figure 4). The model visualizes the relationship between two macro phenomena (A and D) and attitudes and actions on the micro level (B and C). The model is based on the macro-micro-macro scheme published initially by Coleman (1990); it is based on the premise that a macro phenomenon should be explained by actions on the micro-level. The DBO theory (Hedström & Ylikoski, 2010) is helpful for interpreting these actions. According to this theory, peoples' actions are influenced and caused by their desires (D), beliefs (B), and opportunities (O) (Hedström, 2006). The adaption of the scheme to evaluative bibliometrics is helpful in integrating the different functions and reasons to cite (which are on the micro level) in a superordinate context of the general norm in science to cite, and the use of bibliometric data for performance measurements. The model reveals that the results of evaluative citation analyses depend on the actions (cite papers one or multiple times in a document or discuss cited papers extensively) being measured.

According to Hedström and Ylikoski (2010), the relationship between macro properties should be understood by explaining the (causal) relationship between micro properties. In order to understand the macro-level relationship, we should obtain a deeper explanatory understanding of the properties at the micro level. According to the model in Figure 4, at the macro level, "the norm in science to cite prior research" leads to actions on the micro-level to realize the norm. However, this can be undertaken in different ways, and depends on the "attitudes of authors" (e.g. authors' belief they should cite the papers of reputable researchers). On the micro level, the various attitudes end in different "actions" (e.g. citing papers multiple times in a document; or discussing cited papers extensively). As the studies in this literature review show, the different actions of the authors can be analyzed in comprehensive studies, since large datasets in machine-readable formats are now available (see section 1.2). It is no longer necessary to remain on the "number of times cited" level in evaluative bibliometrics, since the different actions of citing authors can be taken into account. The aggregation of different citation actions (see section 3) leads to varying results on evaluated units: they will be different if citations are only counted when papers are mentioned multiple times in a paper, or if citations of papers are only summed up which are discussed in more detail in the citing papers. Thus, the phenomenon on the macro level – the results of evaluative studies on entities (researchers, institutions, countries etc.) – depends on what is counted. The model in Figure 4 also reveals that the results regarding the evaluated units depend on the "attitudes" of the citing authors. Thus, in the interpretation of the corresponding results, the possible attributes should be considered.

Hedström and Ylikoski (2010, p. 59) note that "individuals' actions typically are oriented toward others, and their relations to others therefore are central when it comes to explaining why they do what they do". Thus, we maintain that citing is not solely the result of a single author's decision, but it is typically oriented toward other people as well. The conceptual model published by Tahamtan and Bornmann (2018a) shows that citing decisions might depend on "author", "document" or "journal features". For instance, citing a certain document might be influenced by the editor of the journal to which a citing paper is submitted.

## 5 Conclusions

The purpose of this paper is to update the review of Bornmann and Daniel (2008), presenting a narrative review of studies on the citing behavior of scientists. The current review covers 41 studies published between 2006 and 2018. Bornmann and Daniel (2008) focused on earlier years. The current review describes the (new) studies on citation content and context analyses as well as the studies that explore the citation motivation of scholars through surveys or interviews. The identification of the functional relationships between citing and cited documents in large databases, such as WoS or Scopus, could be used to develop and improve citation indexing and information retrieval tools. Di Marco et al. (2006) proposed that a citation indexing tool would provide information about the relationship between the cited and citing document. They mentioned that such an indexing tool could use "automated citation classification" to determine the functions of citations. Di Marco et al. (2006) proposed that citation functions could be added to information retrieval systems as a new feature to enhance their retrieval capabilities. This feature also could be added to citation indices through the use of tools that are able to automatically classify citations into one or more functional categories (Di Marco et al., 2006; Wang et al., 2012). Information retrieval search engines and databases could use the context of citations to find relevant papers and to show how these papers are related to each other (Aljaber, Martinez, Stokes, & Bailey, 2011; Aljaber, Stokes, Bailey, & Pei, 2010; Di Marco et al., 2006; Lakshmanan & Ramanathan, 2019). In this regard, Ritchie et al. (2008) showed that information retrieval performance increased when more terms from the citation context were indexed. Aljaber et al. (2010) demonstrated that using the citation context terms improved the performance of document clustering, which is used in information retrieval in order to identify relevant information. Dabrowska and Larsen (2015) indicated that relatively short citation contexts improved information retrieval performance. Liu et al. (2014) designed a retrieval system based on the topic words that were extracted from the citation contexts of papers and demonstrated that the system was more successful than Google Scholar and PubMed in finding relevant references.

Citation errors seem to be common among cited references and citation context, which might influence the results of citation context studies (Anderson, 2006). However, in large-scale studies (on a high aggregation level) it will (probably) not have a large influence on results. One can assume that the errors are not systematically distributed. Anderson (2006) found several cases in which the date and title of Karl Weick's book and the author's name were not properly cited. Hammarfelt (2011) revealed some errors in the references citing Walter Benjamin's book, such as variations in the author's name. Wright and Armstrong (2008, p. 125) refer to a phenomenon called "faulty citations" which includes the "omissions of relevant papers, incorrect references, and quotation errors that misreport findings". To address these citation issues, Wright and Armstrong (2008) analyzed citations to the highly-cited paper of Armstrong and Overton (1977) and found that 49 of the 50 studies had reported the findings of Armstrong and Overton (1977) incorrectly. 7.7% incorrect cited references were found for this paper, including 36 variations of the reference. Wright and Armstrong (2008) noted that faulty citations mainly appear because authors fail to read the original paper or do no fully comprehend it. This is in line with the study by Ramos et al. (2012), who noted that cited documents are not read at all by the vast majority of citing authors, or are read carelessly. Ramos et al. (2012, p. 717) found several instances of "'incorrect attribution', which refers to attributing information to the wrong author". However, Thornley et al. (2015) showed contrary results, referring to the argument that researchers carefully read the documents they cite.

Author biographies

Iman Tahamtan is a Ph.D. Student and Graduate Teaching Associate in the School of Information Sciences, College of Communication and Information, at the University of Tennessee in Knoxville (USA). He is the Communications Officer at SIG Health Informatics at the Association for Information Science and Technology (ASIS&T). His current research interests include scholarly communication, bibliometrics, and health information. He teaches information technology courses for undergraduate students. He has received several awards in recent years, including the 2018 Graduate Student Research Award in the College of Communication and Information at the University of Tennessee, Outstanding Achievements in Research Contribution to the University of Tennessee in 2017, the 2019 CHIIR'19/SIGIR Student Travel Grants, Glasgow, Scotland, and the 2018 ASIS&T SIG-USE Student Travel Award, Vancouver. Canada.

Lutz Bornmann is a habilitated sociologist of science and works at the Division for Science and Innovation Studies in the Administrative Headquarters of the Max Planck Society in Munich (Germany). His research interests include research evaluation, peer review, bibliometrics, and altmetrics. He is member of the editorial board of *Quantitative Science Studies* (MIT Press), *PLOS ONE* (Public Library of Science), and *Scientometrics* (Springer) as well as advisory editorial board member of *EMBO Reports* (Nature Publishing group). Clarivate Analytics (http://highlycited.com) lists him among the most-highly-cited researchers worldwide over the last ten years (since the first release of this service in 2014). He is recipient of the *Derek de Solla Price Memorial Medal* in 2019.